\documentclass[12pt,letterpaper]{article}
\usepackage[usenames, dvipsnames]{xcolor}
\usepackage{jcapmod}

\usepackage{tocloft}
\usepackage[]{todonotes}
\usepackage{verbatim}
\usepackage{mathrsfs}
\usepackage{cleveref}
\usepackage[shortlabels]{enumitem}
\usepackage{pgf}
\usepackage{tikz-cd}
\usetikzlibrary{shapes,arrows}
\usetikzlibrary{calc}
\usepackage{pgfplots}
\pgfplotsset{compat=1.12}
\usepackage{tikz-3dplot}

\usepackage{amsthm}

\usepackage{tikz}
\usepackage{setspace,caption}
\usepackage{lipsum}
\usetikzlibrary{matrix,arrows,calc}
\makeatletter
\newsavebox\myboxA
\newsavebox\myboxB
\newlength\mylenA

\definecolor{cornellRed}{HTML}{B31B1B}
\definecolor{cornellBlue}{HTML}{0068AC}
\definecolor{cornellGreen}{HTML}{6EB43F}

\usepackage{bbm} 					
\usepackage{slashed} 				
\usepackage{graphicx}				
\usepackage{subcaption}			
\usepackage{psfrag}				
\usepackage{tensor}				
\usepackage{fouridx}				
\usepackage{bm}					
\usepackage{mdframed}				
\usepackage{multirow}				
\usepackage{soul}					
\usepackage{bbold}				
\usepackage{multicol}				
\usepackage{tikz-cd}
\usepackage{rotating}

\usetikzlibrary{arrows}

\tikzset{
commutative diagrams/.cd,
arrow style=tikz,
diagrams={>=latex}}

\usepackage{amsmath}
\usepackage{amssymb}
\usepackage{amsfonts}
\usepackage{mathtools}

\usepackage{feynmf}
\usepackage{marvosym}

\usepackage{import}

\newcommand*\xoverline[2][0.75]{%
    \sbox{\myboxA}{$\m@th#2$}%
    \setbox\myboxB\null
    \ht\myboxB=\ht\myboxA%
    \dp\myboxB=\dp\myboxA%
    \wd\myboxB=#1\wd\myboxA
    \sbox\myboxB{$\m@th\overline{\copy\myboxB}$}
    \setlength\mylenA{\the\wd\myboxA}
    \addtolength\mylenA{-\the\wd\myboxB}%
    \ifdim\wd\myboxB<\wd\myboxA%
       \rlap{\hskip 0.5\mylenA\usebox\myboxB}{\usebox\myboxA}%
    \else
        \hskip -0.5\mylenA\rlap{\usebox\myboxA}{\hskip 0.5\mylenA\usebox\myboxB}%
    \fi}
\makeatother

\newcommand{\im}{\,\mathrm{Im}\,}
\newcommand{\re}{\,\mathrm{Re}\,}

\definecolor{cobalt}{RGB}{44, 98, 120}
\definecolor{celadon}{rgb}{0.67, 0.88, 0.69}
\definecolor{dm}{cmyk}{.20, 0, .30, 0}
\definecolor{burgundy}{rgb}{0.5, 0.0, 0.13}
\definecolor{plotBlue}{RGB}{94, 130, 181}

\hypersetup{
  colorlinks,
  citecolor=Violet,
  linkcolor=cobalt,
  urlcolor=Blue}

\DeclareSymbolFontAlphabet{\mathbb}{AMSb}

\newif\iffastcompile

\fastcompilefalse

\iffastcompile
\newcommand{\mk}[1]{}
\else
\newcommand{\mk}[1]{\todo[color=burgundy!30, size=\scriptsize, bordercolor=burgundy!30]{MK: #1}}
\fi

\ProvideTextCommandDefault{\Dbar}{%
\leavevmode\lower.5ex\rlap{\hskip-.07em\accent"16}D%
}

\usepackage{environ}
\usepackage{changepage}

\begin{document}
	\newcommand{\main}{.}
\begin{titlepage}

\setcounter{page}{1} \baselineskip=15.5pt \thispagestyle{empty}
\setcounter{tocdepth}{2}
	{\hfill \small MIT-CTP/5552}
\bigskip\

\vspace{1cm}
\begin{center}
{\large \bfseries On one-loop corrected dilaton action in string theory}
\end{center}

\vspace{0.55cm}

\begin{center}
\scalebox{0.95}[0.95]{{\fontsize{14}{30}\selectfont Manki Kim$^{a}$\vspace{0.25cm}}}

\end{center}

\begin{center}

\vspace{0.15 cm}
{\fontsize{11}{30}
\textsl{$^{a}$Center for Theoretical Physics, Department of Physics, Massachusetts Institute of Technology, Cambridge, MA 02139}}\\
\vspace{0.25cm}

\vskip .5cm
\end{center}

\vspace{0.8cm}
\noindent

\vspace{1.1cm}
This manuscript concerns string one-loop corrections to the K\"{a}hler potential in 4d $\mathcal{N}=1$ vacua of string theories, and it largely consists of two parts. In the first part, we compute the string one-loop correction to the dilaton kinetic term in heterotic string theories, and we show that the dilaton kinetic term is not renormalized at one-loop. This result agrees with the background field method studied by Kiritsis and Kounnas. After reviewing the well-known result on the string one-loop correction to the K\"{a}hler potential in heterotic string theories, we explain how this result can be reconciled with the known result on the K\"{a}hler potential in heterotic string theory. In the second part, we study the dilaton dependence of the one-loop corrected K\"{a}hler potential in type II string theories. To do so, we compute the string one-loop correction to the kinetic action of the 4d-dilaton in type II string theories compactified on orientifolds of Calabi-Yau threefolds. We find that the string one-loop corrected 4d dilaton kinetic term is determined by the Witten index and the new supersymmetric index of the string worldsheet CFT.

\vspace{3.1cm}

\noindent\today

\end{titlepage}
\tableofcontents\newpage

\section{Introduction}
The quest for exhibiting realistic vacua of string theory remains one of the most important open problems in string phenomenology. The challenge is in part due to the lack of computational tools to fully characterize the effective action of the low-energy effective theory of string theory. With our current understanding of string theory, studying string compactifications in which supersymmetry is explicitly broken at high energy seems practically intangible as much of the computational tools rely on supersymmetry. On the other hand, string compactifications with minimal supersymmetry at high energy provide an interesting arena. Because of the presence of minimal supersymmetry, there is hope that one can characterize the low-energy supergravity rather precisely, and the low-energy supergravity does not seem utterly unrealistic with the assumption that spontaneous supersymmetry breaking at low-energy can be achieved reliably.\footnote{Low-energy supersymmetry breaking here means low-energy compared to the string scale. To find realistic vacua, the supersymmetry breaking scale should be higher than TeV.} Perhaps the most well studied proposals to construct de Sitter vacua in string theory are the KKLT proposal and the variants thereof \cite{Kachru:2003aw,Balasubramanian:2005zx}. Recently there have been flourishing activities investigating the vacuum structure of 4d $\mathcal{N}=1$ compactifications \cite{Demirtas:2019sip,Carta:2020ohw,Alvarez-Garcia:2020pxd,Scalisi:2020jal,Demirtas:2020ffz,Gao:2020xqh,Grimm:2020cda,Bena:2021wyr,Honma:2021klo,Marchesano:2021gyv,Cicoli:2021dhg,Demirtas:2021nlu,Bastian:2021hpc,Carta:2021kpk,Carta:2021lqg,Bena:2022cwb,Gendler:2022qof,Gendler:2022ztv,Cicoli:2022vny,Cribiori:2021djm,Krippendorf:2022gcl,Shiu:2022oti,Schreyer:2022len,Hebecker:2022zme,Moritz:2023jdb}. In this spirit, we shall study the 4d $\mathcal{N}=1$ supergravity theories derived from string compactifications.

The vacuum structure of the 4d $\mathcal{N}=1$ low-energy effective theory of string theory is characterized by the K\"{a}hler potential, the superpotential, and the holomorphic gauge coupling of the low-energy supergravity. The superpotential and the holomorphic gauge couplings are holomorphic functions of moduli. The holomorphicity of such terms allows us to compute them very precisely. The K\"{a}hler potential, on the other hand, is a real function of moduli. As a result, the K\"{a}hler potential does not enjoy any non-renormalization properties unlike the superpotential and the holomorphic gauge coupling. Intricacies of computing the K\"{a}hler potential precisely has been one of the main challenges for finding cosmological solutions of string theories.\footnote{For recent reviews on cosmological vacua of string theory, see \cite{Hebecker:2020aqr,Flauger:2022hie,Cicoli:2023opf,VanRiet:2023pnx}.}

Na\"{i}vely, one expects that to compute the K\"{a}hler potential one must know the string spectrum in gross detail. This reasonable expectation is one of many reasons why until recently the loop corrections to the K\"{a}hler potential were only computed in toroidal orientifold compactifications. Furthermore, even with the knowledge of the full string spectrum in toroidal orientifolds, one must perform extremely daunting computations to obtain the open string one-loop correction to the K\"{a}hler potential \cite{Antoniadis:1997eg,Berg:2005ja,Berg:2014ama,Haack:2015pbv,Haack:2018ufg}.\footnote{In the context of 4d $\mathcal{N}=2$ compactifications, by pioneering works it is known that the loop corrections to the K\"{a}hler potential are determined by the Witten index of the Calabi-Yau CFT in type II string theories \cite{Kiritsis:1994ta,Antoniadis:1997eg,Strominger:1997eb,Antoniadis:2003sw}, and the K3 BPS indices in heterotic string theory and type I string theory \cite{Harvey:1995fq,Antoniadis:1996vw}. See also \cite{Kiritsis:2001bc,Antoniadis:2002tr} for interesting works on corrections to the universal sector in the context of the localized gravity around D-branes.} Therefore, one can argue that the computation of the string one-loop correction to the K\"{a}hler potential in a generic Calabi-Yau orientifold on the surface seems hopeless.\footnote{Despite the difficulties, there have been significant efforts to understand corrections to the K\"{a}hler potential in Calabi-Yau orientifold compactifications \cite{Giddings:2005ff,Berg:2007wt,Grimm:2007xm,Cicoli:2007xp,Douglas:2008jx,Cicoli:2008va,Garcia-Etxebarria:2012bio,Junghans:2014zla,Martucci:2014ska,Martucci:2016pzt,Antoniadis:2019rkh,Cicoli:2021rub,Gao:2022uop,Lust:2022xoq}.} 

 
But, it is worth emphasizing that earlier works on the loop corrections have shown that various one-loop corrections to the Kahler potential are actually rather special than one might have na\"{i}vely expected, even in minimally supersymmetric string compactifications. In the context of heterotic string theory, using the background field methods, it was shown that in supersymmetric compactifications of the Einstein-Hilbert action, the axion kinetic term, and the dilaton kinetic term are shown to be not renormalized in string perturbation theory \cite{Kiritsis:1994ta}. Furthermore, in \cite{Antoniadis:1992rq,Antoniadis:1992pm,Kaplunovsky:1992vs,Kaplunovsky:1995jw} the one-loop correction to the K\"{a}hler potential and the gauge threshold correction were shown to be related to the new-supersymmetric index \cite{Cecotti:1992qh}.

Continuing the surprise, in recent work \cite{Kim:2023sfs}, it was realized that the string one-loop correction to the Einstein-Hilbert action in string-frame is determined by \emph{the new supersymmetric index}, which was studied in \cite{Cecotti:1992qh},
\begin{equation}
\text{Tr}_R\left( (-1)^FFq^{L_0-\frac{c}{24}}\right)\,.
\end{equation}
For example, the annulus contribution to the string one-loop correction to the Einstein-Hilbert action in string-frame is given as
\begin{equation}
\delta E_A=\frac{1}{2^{8}\pi^2}\int_0^\infty \frac{dt}{2t^2}\left[\text{Tr}_R\left((-1)^{F-\frac{3}{2}}Fq^{L_0-\frac{3}{8}}\right)^{open}_{int}+\frac{3}{2}(n_A^+-n_A^-)\right]\,.
\end{equation}
As the new susy index is sensitive only to the F-terms of the underlying theory, it might be possible to compute $\delta E_A$ explicitly in explicit models of Calabi-Yau compactifications. For example, it is well known that for the integrable models the new susy index can be computed using the thermodynamic Bethe ansatz \cite{Zamolodchikov:1989cf}. Furthermore, it is well known that the integral
\begin{equation}
\int\frac{dt}{2t}\text{Tr}_R\left( (-1)^FFq^{L_0-\frac{c}{24}}\right)\,,
\end{equation}
can be computed exactly using the $tt^*$ equations \cite{Bershadsky:1993cx,Walcher:2007tp,Walcher:2007qp}.

In this draft, we continue the development of computational tools to compute string one-loop correction to the K\"{a}hler potential for type II string compactifications on Calabi-Yau orientifolds. More precisely, we shall compute the string one-loop correction to the kinetic term of the 4d-dilaton which is defined as in large volume limit
\begin{equation}
e^{-2\phi_4}:= \mathcal{V} e^{-2\phi}\,,
\end{equation}
where $\mathcal{V}$ is string-frame Calabi-Yau volume, and $\phi$ is the 10d-dilaton. Because the 4d-dilaton depends both on the Calabi-Yau volume and the axio-dilaton, understanding the coupling of the 4d-dilaton is absolutely crucial to fully characterize the moduli kinetic term. Furthermore, because the 4d-dilaton is universal, the computation involving the 4d-dilaton is much easier compared to the scattering amplitudes involving the exactly marginal deformations of the Calabi-Yau. As a result, we will be able to perform $\alpha'$ exact computation at string one-loop. 

To translate the string loop amplitude to the effective action, one must carefully check which frame one is using. At string tree-level, it is well known that the two-point function of the commonly used 4d-dilaton vertex operator $V_D$ computes the 4d-dilaton kinetic term in \emph{Einstein-frame} not \emph{string-frame} \cite{Blumenhagen:2013fgp,Haack:2018ufg}. Because the dilaton vertex operator corresponds to the variation of the 4d-dilaton in Einstein-frame at string tree-level, the kinetic term of the 4d-dilaton computed by the dilaton two-point function $\langle\langle V_D V_D\rangle\rangle,$ will neither be in Einstein-frame or string-frame at string one-loop or beyond. We should therefore define a new frame, the vertex-frame in which the 4d-dilaton kinetic term is directly related to the dilaton two-point function of the string amplitude. This is due to how Einstein-frame is commonly defined to be the frame in which the Einstein-Hilbert action is canonically normalized. In this draft, we will show the clear distinction between these three different frames. 

We will find that in type II string theories the string one-loop correction to the 4d dilaton kinetic term in the \emph{vertex-frame} from the even spin structures yields
\begin{equation}\label{eqn:intro main1}
\boxed{\mathfrak{G}_{\sigma,\phi}^{(1),(even)}=-\frac{c_\sigma}{2^3\pi^2} \int_0^\infty \frac{dt}{2t^3}\tau_{2\sigma}\left[\text{Tr}_R\left((-1)^{F-\frac{3}{2}}Fq^{L_0-\frac{3}{8}}\right)^{\sigma}_{int}+\frac{3}{2}(n_\sigma^+-n_\sigma^-)\right]\,.}
\end{equation}
and
\begin{equation}
\mathfrak{G}_{\phi}^{(1),(even)}=\frac{1}{2}\left(\mathfrak{G}_{\mathcal{A},\phi}^{(1),(even)}+\mathfrak{G}_{\mathcal{M},\phi}^{(1),(even)}+\mathfrak{G}_{\mathcal{K},\phi}^{(1),(even)}\right)\,,
\end{equation}
where we define
\begin{equation}
\mathfrak{G}^{(1)}_{\phi_4}:= \delta G_{\phi_4}-3\delta E\,,
\end{equation}
see for example \eqref{eqn:Einstein frame effective action0}. Similarly, we find that the odd spin structure contribution is given by \eqref{eqn:main result}
\begin{equation}\label{eqn:intro main2}
\boxed{\mathfrak{G}_{\phi_4}^{(1),(odd)}=\frac{1}{2}(\mathfrak{G}_{\phi_4,\mathcal{T}}^{(1),(odd)}+\mathfrak{G}_{\phi_4,\mathcal{T}}^{(1),(odd)})=\pm\frac{1}{2^5 \cdot 3\cdot \pi} (\chi+\chi_f)\,.}
\end{equation}
Combining \eqref{eqn:intro main1} and \eqref{eqn:intro main2}, we arrive at the main result of this paper
\begin{equation}
\boxed{\mathfrak{G}_{\phi}^{(1)}=\mathfrak{G}_{\phi}^{(1),(even)}+\mathfrak{G}_{\phi_4}^{(1),(odd)}\,.}
\end{equation}


This paper is organized as follows. In \S\ref{sec:effective action}, we study the effective action at string one-loop. We shall study how to translate the string scattering into the loop correction to the effective action. Furthermore, we will explain which terms need to be computed in future work to complete the computation of the one-loop corrected K\"{a}hler potential. In \S\ref{sec:het}, we study the string one-loop correction to the effective action of heterotic string theory compactified on Calabi-Yau threefolds. We will find that the 4d-dilaton kinetic term is not renormalized at string one-loop. We will review the string one-loop correction to the K\"{a}hler potential, and its relation to the Green-Schwarz term in the linear multiplet formalism. After the review, we will explain how the absence of the one-loop correction to the 4d dilaton kinetic term is in agreement with the known results on the loop correction to the  K\"{a}hler potential. In \S\ref{sec:type II}, we will compute the string one-loop correction to the 4d-dilaton kinetic action in type II string theories. We will find that the string one-loop correction to the 4d dilaton kinetic term is determined by the Witten index and the new supersymmetric index of the internal CFT. In \S\ref{sec:conclusion}, we conclude. In \S\ref{sec:Green function}, we collect the conventions used in this draft. In \S\ref{app:cancellation}, we compute the genuine contractions of the eight fermion correlator.

\emph{Note added: Shortly after the submission of the draft, it was found that even at the four-fermion level naive $\mathcal{O}(\delta^2)$ terms can in fact yield $\mathcal{O}(\delta)$ contribution due to the vertex collision. Including this effect invalidates the wrong claim of the v1.
that the open string contributions to the 4d dilaton kinetic term in the vertex frame cancel. Corrected result is still written in terms of the Witten index and the new supersymmetric index of the internal CFT. }
\section{String one-loop corrected effective action}\label{sec:effective action}
In this section, we will study two things. First, how to relate the dilaton two-point function at string one-loop to string one-loop correction to the effective action in string-frame. Second, we will explain how to obtain the kinetic term of moduli fields at string one-loop in Einstein-frame, with an emphasis on which terms have been computed in the literature so far and what needs to be further computed. In \S\ref{eqn:eff action}, although most of the discussion is easily generalizable to other string theories, we will focus on the effective action of type IIB string theory compactified on an O3/O7 orientifold of a Calabi-Yau threefold. We summarize different frames used in this paper in table \ref{table1}.

\subsection{One-loop correction to the dilaton kinetic term}
In this section, we shall explain how to relate the string one-loop correction to the kinetic term of the 4d dilaton to the scattering amplitudes of two dilatons. Let us start by considering the one-loop correction to the 4d-dilaton kinetic term
\begin{equation}\label{eqn:dilaton kinetic}
S_{st}^{(1)}\supset-\frac{1}{\kappa_4^2}\int d^4x\sqrt{-g} G^{(1)}_{\phi_4}(\partial \phi_4)^2\,,
\end{equation}
and the one-loop correction to the Einstein-Hilbert action in \emph{string-frame}
\begin{equation}
S_{st}^{(1)}\supset\frac{1}{2\kappa_4^2}\int d^4x\sqrt{-g}\delta E R\,,
\end{equation}
where we normalize the Einstein-Hilbert action at the string tree level as
\begin{equation}
\frac{1}{2\kappa_4^2}\int d^4x e^{- 2\phi_4} R\,.
\end{equation}

We shall determine the kinetic term \eqref{eqn:dilaton kinetic} by comparing the action to the one-loop scattering amplitude. Let us consider the dilaton two-point function summed over Riemann surfaces with or without boundaries of genus 1
\begin{equation}\label{eqn:dilaton two point function}
Z=\sum_R Z_R=\sum_R \langle\langle V_D (p_1,\epsilon_1) V_D(p_2,\epsilon_2)\rangle\rangle_R\,,
\end{equation}
where $V_D$ is the vertex operator that corresponds to the dilaton state. We will study the explicit form of the dilaton vertex operator later in this draft. The amplitude \eqref{eqn:dilaton two point function} formally vanishes on-shell due to the momentum conservation condition $p_1\cdot p_2=0.$ To extract the one-loop renormalization of the kinetic term, one can let the momentum $\xi:=p_1+p_2$ vary on a complex plane without imposing the momentum conservation condition \cite{Minahan:1987ha}. This amounts to not integrating over the bosonic zero modes of the strings.\footnote{On a related note, recently it was shown that the string theoretic two-point function at string tree-level is equivalent to field theoretic two-point function \cite{Erbin:2019uiz}.} Unlike evaluating generic off-shell amplitudes, we can still reliably use the worldsheet CFT, as the only off-shellness is coming from $\xi\neq0.$ Phrased differently, every state that contributes to the amplitude is on-shell. Then, in the small momentum expansion, we formally have an expansion
\begin{equation}
Z= AV_4 g_c^2 p_1\cdot p_2+\mathcal{O}(p^4)\,,
\end{equation}
where $V_4$ is a properly regulated volume of the non-compact dimensions. 

As we shall see later in this draft, we will define the dilaton vertex operator such that at string tree-level the dilaton vertex operator corresponds to the 4d dilaton in \emph{Einstein-frame}, c.f. section 16 of \cite{Blumenhagen:2013fgp}.\footnote{We thank Michael Haack for explaining this to us.} We identify the variation of the 4d dilaton in \emph{Einstein-frame} as
\begin{equation}
\delta \phi_4 =-2\pi g_c \int \frac{d^4k}{(2\pi)^4} e^{ip\cdot x}\,,
\end{equation}
with the dilaton vertex operator $V_D.$ Equivalently, in \emph{string-frame}, we can identify $V_D$ with the variation of the 4d dilaton $\delta \phi_4,$ and the metric
\begin{equation}
\delta g_{\mu\nu}=2\eta_{\mu\nu}\delta \phi_4\,.
\end{equation}
As we shall find in the next section, the one-loop correction to the kinetic term of the 4d-dilaton in the vertex-frame is
\begin{equation}
-\frac{1}{\kappa_4^2}\int d^4 x\sqrt{-g} (G_{\phi_4}^{(1)}-3\delta E)(\partial \delta\phi_4)^2\,,
\end{equation}
therefore we find 
\begin{equation}\label{eqn:identification}
\mathfrak{G}_{\phi_4}^{(1)}:=(G_{\phi_4}^{(1)}-3\delta E)=\frac{\kappa_4^2}{8\pi^2}A\,.
\end{equation}
This relation will be heavily used to determine the one-loop corrected kinetic action.

\subsection{Effective action for O3/O7 orientifolds}\label{eqn:eff action}
In this section, we shall study the effective action for type IIB compactifications on O3/O7 orientifolds of Calabi-Yau threefolds. The goal of this section is twofold. First, we shall explain how to translate the computation done in the previous sections to the one-loop corrected action in Einstein-frame. Second, by doing so, we will naturally set up a stage to explain what computations must be done to complete the computation of the one-loop corrected K\"{a}hler potential. Most of the discussion of this section is sufficiently general and hence can be generalized to other orientifold compactifications. To simplify the discussion, only in this section, we will assume $h^{1,1}_-(X_3)=h^{2,1}_+(X_3)=0.$ We stress that the computation carried out in \S\ref{sec:type II} applies to any Calabi-Yau orientifolds.

\subsubsection{String tree level}
Let us first start by reviewing the effective action at string-tree level \cite{Becker:2002nn,Grimm:2005fa}. We shall focus on graviton, dilaton, K\"{a}hler moduli, and complex structure moduli. The full action including the rest of the field can be recovered by demanding supersymmetry. The effective action at the string-tree level in string-frame is then written as
\begin{equation}\label{eqn:tree-level effective action0}
S^{(0)}\supset \frac{1}{2\kappa_4^2}\int d^4 x \sqrt{-g} e^{-2\phi_4}\left[R+4 (\partial \phi_4)^2 -2 G_{a\bar{b}}\partial t^a\partial \bar{t}^{\bar{b}}-2G_{\alpha\bar{\beta}}\partial z^\alpha\partial \bar{z}^{\bar{\beta}}\right]\,,
\end{equation}
where we define in large volume limit
\begin{equation}
e^{-2\phi_4}:= e^{-2\phi} \left(\mathcal{V}-\frac{\zeta(3)\chi}{4(2\pi)^3}\right)\,,
\end{equation}
\begin{equation}
G= -\log \left(\mathcal{V}-\frac{\zeta(3)\chi}{4(2\pi)^3}\right)-\log \left(i\int \Omega\wedge\overline{\Omega}\right)\,,
\end{equation}
and
\begin{equation}
G_{a\bar{b}}:=\partial_{t^a}\partial_{\bar{t}^{\bar{b}}} G\,,\quad G_{\alpha\bar{\beta}}:=\partial_{z^\alpha}\partial_{\bar{z}^{\bar{\beta}}} G\,.
\end{equation}
Note here that $t^a$ is a complexified string-frame volume of a curve $C_a$, $\mathcal{V}$ is volume of the Calabi-Yau threefold in string-frame, and $z^a$ is a complex structure modulus. The kinetic term $G_{a\bar{b}}$ and $G_{\alpha\bar{\beta}}$ can be determined via Zamolochikov metric of the Calabi-Yau CFT \cite{Zamolodchikov:1986gt}. 

At string tree-level, the vertex-frame coincides with Einstein-frame, up to rescaling by a factor of $e^{2\phi_4^*}$ of the dilaton kinetic term. Let us study the vertex-frame. We shall define the vertex frame as a frame determined by the variations of the following fields
\begin{equation}\label{eqn:vertex var00}
g_{\mu\nu}=\eta_{\mu\nu}+2 \eta_{\mu\nu} \delta \phi_4\,,
\end{equation}
and
\begin{equation}\label{eqn:vertex var10}
\phi_4=\phi_4^*+\delta\phi_4\,,
\end{equation}
in \emph{string-frame}. Because we are doing string perturbation theory around the flat background, the zero-th order metric is set to be 4d Lorentzian metric. Similarly, $\phi_4^*$ is the background value of the 4d dilaton. Then, using the identities
\begin{equation}\label{eqn:vertex EH0}
\frac{1}{2\kappa_4^2}\int d^4 x \sqrt{-g} e^{-2\phi_4}R=\frac{1}{2\kappa_4^2}\int d^4x\sqrt{-\eta} e^{-2\phi_4^*} [-6(\partial \delta\phi_4)^2]\,,
\end{equation}
we can rewrite the effective for the 4d-dilaton in the vertex-frame as
\begin{equation}\label{eqn:vertex frame dilaton kinetic tree-level}
S^{(0)}\supset -\frac{1}{\kappa_4^2}\int d^4 x\sqrt{-\eta} e^{-2\phi_4^*} (\partial\delta\phi_4)^2\,.
\end{equation}

Now, we shall study Einstein-frame. To obtain the moduli kinetic term in Einstein-frame, we rescale the metric as
\begin{equation}
g\mapsto e^{2\phi_4} g\,,
\end{equation}
which maps the Einstein-Hilbert term in string-frame to
\begin{equation}
\frac{1}{2\kappa_4^2}\int d^4 x \sqrt{-g} \left[ R-6 (\partial \phi_4)^2\right]\,.
\end{equation}
As a result, we obtain the well known form of the kinetic term in Einstein-frame \cite{Becker:2002nn,Grimm:2005fa}
\begin{equation}\label{eqn:tree level action IIB}
-\frac{1}{\kappa_4^2} \int d^4 x \sqrt{-g} \left[(\partial \phi_4)^2+G_{a\bar{b}}\partial t^a\partial \bar{t}^{\bar{b}}+G_{\alpha\bar{\beta}}\partial z^\alpha\partial \bar{z}^{\bar{\beta}}\right]\,.
\end{equation}
As we advertised, \eqref{eqn:vertex frame dilaton kinetic tree-level} is equivalent to the 4d dilaton kinetic action in Einstein-frame under the rescaling by
\begin{equation}
g\mapsto g e^{-2\phi_4^*}\,.
\end{equation} 

Although the action \eqref{eqn:tree level action IIB} is complete, the action is not written in terms of the canonical chiral fields. We shall rewrite \eqref{eqn:tree level action IIB} in terms of the following chiral multiplets 
\begin{equation}
\tau=C_0+ie^{-\phi}\,,\quad T_a=\frac{e^{-\phi}}{2}\mathcal{K}_{abc}t^bt^c+i\int_{D^a} C_4+\dots\,,
\end{equation} 
where $D^a$ is a divisor dual to a curve $C_a$, and $\dots$ represents non-perturbative corrections which we neglect in this work. By using relations
\begin{equation}
-2\phi_4= -\frac{1}{2}\phi+\log \left(\mathcal{V}_E-g_s^{3/2}\frac{\zeta(3)\chi }{4(2\pi)^3}\right)\,,
\end{equation}
\begin{equation}
G=-\log \left(\mathcal{V}_E-g_s^{3/2}\frac{\zeta(3)\chi }{4(2\pi)^3}\right)+\frac{3}{2}\log(\tau-\bar{\tau})-\log \left(i\int \Omega\wedge\overline{\Omega}\right)\,,
\end{equation}
we recover the standard form of the K\"{a}hler potential \cite{Becker:2002nn}
\begin{equation}
\mathcal{K}=-\log (\tau-\bar{\tau})-2 \log \left(\mathcal{V}_E-g_s^{3/2}\frac{\zeta(3)\chi }{4(2\pi)^3}\right)-\log \left(i\int \Omega\wedge\overline{\Omega}\right)\,,
\end{equation}
and the kinetic term
\begin{equation}\label{eqn:kinetic type IIB einstein0}
S\supset-\frac{1}{\kappa_4^2}\int d^4 x\sqrt{-g}\, \mathcal{K}_{\varphi^a\bar{\varphi}^{\bar{b}}}\partial \varphi^a\partial \bar{\varphi}^{\bar{b}}\,.
\end{equation}

\subsubsection{String one-loop level}
We shall now add the string one-loop corrections to the effective action in string-frame
\begin{align}
S^{(1)}=&\frac{1}{2\kappa_4^2}\int d^4x \biggl[\delta E R-2\delta G_{\phi_4} (\partial \phi_4)^2- 2\delta G_{a\bar{b}}\partial t^a \partial \bar{t}^{\bar{b}}-2\delta G_{\alpha\bar{\beta}}\partial z^\alpha\partial \bar{z}^{\bar{\beta}}\nonumber\\
&\qquad\qquad\qquad-\delta G_{a\bar{\beta}}\partial t^a\partial \bar{z}^{\bar{\beta}}-\delta G_{\alpha\bar{b}}\partial z^\alpha\partial \bar{t}^{\bar{b}}\biggr]\,.
\end{align} 
$\delta E$ term was recently computed in \cite{Kim:2023sfs}. We shall compute $\delta G_{\phi_4}$ or more precisely, $\delta G_{\phi_4}-3\delta E,$ in this paper. The rest of the terms are left for future work. 

Let us first determine the effective action that is computed by the string amplitude. We shall focus on the Einstein-Hilbert action and the 4d dilaton kinetic term. We again recall that the dilaton vertex operator corresponds to the variations
\begin{equation}\label{eqn:vertex var1}
g_{\mu\nu}=\eta_{\mu\nu}+2\eta_{\mu\nu}\delta\phi_4\,,
\end{equation}
and
\begin{equation}\label{eqn:vertex var2}
\phi_4=\phi_4^*+\delta\phi_4\,.
\end{equation}
The corresponding variation of the one-loop corrected Einstein-Hilbert action is then
\begin{align}\label{eqn:vertex EH}
\frac{1}{2\kappa_4^2}\int d^4x\sqrt{-g}\, \delta E\, R=&\frac{1}{2\kappa_4^2}\int d^4x\sqrt{-\eta} \left[\delta E(1+2\delta\phi_4)(-6\partial^2\delta\phi_4-6(\partial\delta\phi_4)^2\right]\,\\
=&\frac{1}{\kappa_4^2}\int d^4x\sqrt{-\eta} \left[ 3\delta E (\partial\delta\phi_4)^2\right]\,,
\end{align}
where we fixed all other moduli to be constants. Note here that the sign of \eqref{eqn:vertex EH} is different from the sign of \eqref{eqn:vertex EH0}. This is because the term that contains $-6\partial^2\delta\phi_4$ is not a total derivative in \eqref{eqn:vertex EH}. Similarly, let us vary the 4d-dilaton kinetic term in string-frame according to \eqref{eqn:vertex var1} and \eqref{eqn:vertex var2}
\begin{align}
-\frac{1}{\kappa_4^2}\int d^4x\sqrt{-g} \delta G_{\phi_4} (\partial \phi_4)^2=-\frac{1}{\kappa_4^2}\int d^4x\sqrt{-\eta} \delta G_{\phi_4}(\partial \delta\phi_4)^2\,.
\end{align}
As a result, we conclude that the dilaton two-point function computed in the string amplitude at one-loop corresponds to the following term
\begin{equation}
-\frac{1}{\kappa_4^2}\int d^4x\sqrt{-\eta} (\delta G_{\phi_4}-3\delta E) (\partial \delta \phi_4)^2\,.
\end{equation}
Therefore, we find that the one-loop corrected 4d dilaton kinetic term in the vertex-frame is
\begin{equation}\label{eqn:vertex dilaton action one loop}
-\frac{1}{\kappa_4^2}\int d^4x\sqrt{-\eta} (e^{-2\phi_4^*}+\delta G_{\phi_4}-3\delta E) (\partial \delta \phi_4)^2\,.
\end{equation}

Let us now study Einstein-frame at string one-loop. Because now the Einstein-Hilbert action in string-frame is written as
\begin{equation}
\frac{1}{2\kappa_4^2}\int d^4x \left( e^{-2\phi_4}+\delta E\right) R\,,
\end{equation}
to find the action in Einstein-frame, in which the Einstein-Hilbert action is canonically normalized, we shall rescale the 4d metric as
\begin{equation}\label{eqn:metric rescaling}
g\mapsto g e^{2\Phi_4}=g\left(e^{-2\phi_4}+\delta E\right)^{-1}\,,
\end{equation}
where we defined $e^{-2\Phi_4}:=e^{-2\phi_4}+\delta E.$

After performing the rescaling \eqref{eqn:metric rescaling}, we obtain the one-loop correction to the effective action in Einstein-frame
\begin{align}\label{eqn:Einstein frame effective action0}
S\supset& -\frac{1}{\kappa_4^2}\int d^4x \sqrt{-g}\biggl[ \left(1+(\delta G_{\phi_4}-3 \delta E-G_{\phi_4}\delta E) e^{2\phi_4}\right) (\partial \phi_4)^2-3\partial_\mu\phi_4\partial^\mu(\delta E)e^{2\phi_4}\nonumber\\
&\qquad\qquad\qquad\qquad+\left((1-\delta E e^{2\phi_4})G_{i\bar{j}}+\delta G_{i\bar{j}} e^{2\phi_4}\right)\partial_\mu \varphi^i\partial^\mu\bar{\varphi}^{\bar{j}}\biggr]\,,
\end{align}
where $G_{\phi_4}=-2,$ and $\varphi$ denotes a generic moduli field that is not the 4d dilaton. As a result, we find that the 4d dilaton kinetic term in Einstein-frame is
\begin{equation}\label{eqn:einstein dilaton action one loop}
-\frac{1}{\kappa_4^2}\int d^4x\sqrt{-g} \left(1+(\delta G_{\phi_4}-\delta E)e^{2\phi_4}\right)(\partial\phi_4)^2\,.
\end{equation}

In order to obtain the kinetic term in the canonical basis of chiral multiplets, we shall redefine the fields such that even cycle volumes are now written in terms of Einstein-frame volume
\begin{equation}
e^{-2\phi_4}= \left( \mathcal{V}_E -\frac{g_s^{-3/2}\zeta(3)\chi}{4(2\pi)^3}\right) g_s^{-1/2}\,,
\end{equation}
and
\begin{equation}
t^a=t^a_E g_s^{1/2}\,.
\end{equation}
To simplify the expression, we define
\begin{equation}
\mathcal{Y}=\mathcal{V}_E -\frac{g_s^{-3/2}\zeta(3)\chi}{4(2\pi)^3}\,.
\end{equation}
For the one-loop correction to the dilaton kinetic term in Einstein-frame, we find
\begin{equation}\label{eqn:dilaton effective action 0}
-\frac{e^{2\phi_4}}{\kappa_4^2}\left[(\delta G_{\phi_4}-\delta E)\mathfrak{Z}^2-\frac{3}{2\im\tau} t^a_s\frac{\partial \delta E}{\partial t_s^a}\mathfrak{Z}+\frac{1}{\im\tau^2}\left(-\delta E G_{a\bar{b}}+\delta G_{a\bar{b}}\right) t_s^a\bar{t}_s^{\bar{b}}\right](\partial\im\tau)^2\,,
\end{equation}
where $t_s$ is string-frame curve volume, and we define
\begin{equation}
\mathfrak{Z}=\frac{1}{4}\left[\frac{1}{\im\tau}-\frac{3\zeta(3)\chi}{4(2\pi)^3\mathcal{Y}}\im\tau^{1/2}\right]\,.
\end{equation}
For the one-loop correction to the kinetic for K\"{a}hler moduli in Einstein-frame, we find
\begin{align}
-\frac{e^{2\phi_4}}{\kappa_4^2}&\biggl[\frac{1}{8} (\delta G_{\phi_4}- \delta E) \mathcal{K}_{T_a}\mathcal{K}_{\bar{T}_{\bar{b}}}+ \left(\delta G_{T_a\bar{T}_{\bar{b}}}-\delta E G_{T_a\bar{T}_{\bar{b}}}\right) -\mathfrak{X}\biggr] \partial T_a\bar{\partial} \bar{T}_{\bar{b}}\,,
\end{align}
where we define
\begin{equation}
\mathfrak{X}_{T_a\bar{T}_{\bar{a}}}:=\frac{3}{4} \left(K_{T_a}\frac{\partial \delta E}{\partial \bar{T}_{\bar{b}}}+K_{\bar{T}_{\bar{b}}} \frac{\partial \delta E}{\partial T_a}\right)\,.
\end{equation}
As one can see, the determination of $\delta G_{\phi_4}$ is crucial to complete the one-loop corrected kinetic term of K\"{a}hler moduli, because $\phi_4$ depends on Einstein-frame volume. Similarly, one can easily find the one-loop correction to the kinetic term of other moduli, including the mixing terms. To do so, one needs to take into account that $\delta E$ and $\delta G_{i\bar{j}}$ can in general depend on all moduli, and their overall dilaton dependence is absent. Lastly, to determine the K\"{a}hler potential at string one-loop, one needs to carefully take into account that the holomorphic coordinates in general can be corrected at string one-loop \cite{Haack:2018ufg}. We leave the complete computation of string one-loop corrected moduli kinetic terms in Einstein-frame for future work. 

\begin{table}
\centering
\begin{tabular}{|c|c|c|c|}
\hline
&String-frame&Vertex-frame&Einstein-frame\\\hline
&&&
\\[-1em]
EH term&$\frac{1}{2\kappa_4^2}(e^{-2\phi_4}+\delta E)R$&N/A& $\frac{1}{2\kappa_4^2} R$\\\hline
&&&
\\[-1em]
Dilaton action&$-\frac{1}{\kappa_4^2} (-2e^{-2\phi_4}+\delta G_{\phi_4})(\partial\phi_4)^2$&\eqref{eqn:vertex dilaton action one loop}&\eqref{eqn:einstein dilaton action one loop}\\\hline
&&&
\\[-1em]
Metric&$g=g^{(s)}$&$g=g^{(s)}=\eta+2\eta \delta\phi_4$&$g= g^{(s)}e^{-2\Phi_4}$\\\hline
\end{tabular}
\caption{Comparisons between different frames used in this paper. The vertex-frame corresponds to the frame in which the effective action is determined by the variation of the fields \eqref{eqn:vertex var00} and \eqref{eqn:vertex var10}. Because the metric variation in the vertex-frame corresponds to the spin 0 field, we have not specified the EH action in the vertex-frame. For this table only, we denote the metric in string-frame by $g^{(s)}.$}\label{table1}
\end{table}
\newpage
\section{Heterotic string theory}\label{sec:het}
In this section, we shall study heterotic string compactification on Calabi-Yau threefolds with the standard embedding. Much of the contents in this section are well known to experts. But, understanding the string loop corrections in the context of heterotic string theory will help us jump-start various discussions in type II string theories.

In order to have a consistent compactification of heterotic string theory on a Calabi-Yau threefold, one must cancel the NS5-brane tadpole. To saturate the tadpole without a spacetime filling NS5-brane in the spectrum, one must turn on a vector bundle of subgroups of heterotic gauge groups such that the second Chern class of the vector bundle equals that of the tangent bundle. The simplest choice is to choose the vector bundle to be the tangent bundle, which has $SU(3)$ holonomy. 

The corresponding worldsheet CFT is then constructed as follows. The worldsheet CFT has $(0,1)$ supersymmetry overall, and the matter part has the central charge $(26,15).$ We decompose the worldsheet CFT into three parts. The non-compact directions are described by free field $(\mathcal{N},\bar{\mathcal{N}})=(0,1),$ $(c,\bar{c})=(4,6)$ CFT. The internal part of the CFT decomposes into the geometry part and the Kac-Moody CFT with $(c,\bar{c})=(16,0).$ Because we are identifying the $SU(3)$ vector bundle of the Kac-Moody CFT with the tangent bundle of the internal geometry, we can identify 6 free fermions of the Kac-Moody CFT with the fermionic degrees of freedom of the left moving part of the Calabi-Yau CFT. This determines that the internal CFT is a direct sum of the Calabi-Yau CFT with $(\mathcal{N},\bar{\mathcal{N}})=(2,2)$ and $(c,\bar{c})=(9,9),$ and the Kac-Moody CFT that corresponds to the unbroken gauge group which has the central charge $(c,\bar{c})=(13,0).$

Heterotic string theory compactified on Calabi-Yau threefolds with the standard embedding has many attractive features. First, even though the target space theory in 4d has minimal supersymmetry, the internal CFT of the string worldsheet enjoys extended $\mathcal{N}=2$ superconformal symmetry. As a result, heterotic string theory presents a unique avenue at which one can utilize $\mathcal{N}=2$ techniques to study $\mathcal{N}=1$ vacua of string theory. Second, at string one-loop, there is only one diagram to consider, which greatly simplifies the computation. Third, the string spectrum of the internal CFT is the only necessary data to compute the string one-loop correction. Because the string spectrum of the internal CFT can be approximated in the large volume limit by eigenvalues of the Laplace-Beltrami operators, in principle with the aid of numerical methods to compute the Calabi-Yau metric, the loop corrections can be numerically obtained. And perhaps most importantly, it is very well known how to compute the string one-loop corrected K\"{a}hler potential in heterotic string theory, even in generic Calabi-Yau compactifications. It is the objective of this section to review the relevant knowledge in the old literature of heterotic string theory compactification, and explain how the vanishing one-loop correction to the dilaton kinetic term, that we compute in this draft, is in agreement with the well known results.

\subsection{One loop corrected kinetic action of the dilaton multiplet}
In this section, we shall compute various two-point functions to study the one-loop corrected kinetic term for the dilaton. As is commonly said in the literature, to compute the one-loop corrected K\"{a}hler metric for moduli fields in Einstein frame, one needs to compute string one-loop corrections to the Einstein-Hilbert action and the kinetic term of the moduli fields, both in string frame. In particular, we will find that the 4d dilaton kinetic term and the Einstein-Hilbert action are not renormalized at one loop. This result agrees with the background field method studied in \cite{Kiritsis:1994ta}. 

Assuming this standard lore, based on the results we will present in this section, one can conclude that at string one-loop the K\"{a}hler potential for the dilaton multiplet is not corrected. This sounds contradictory because it is very well known in the literature that the one-loop correction to the K\"{a}hler potential in heterotic string theory depends on the dilaton multiplet, hence the K\"{a}hler potential for the dilaton multiplet is indeed corrected at string one-loop. Although these two statements seem incompatible, we will see that those two statements are not in contradiction, because in the first statement, the dilaton is in linear multiplets, whereas in the second statement, the dilaton is in chiral multiplets. We will come back to this issue in the next section.

We shall first compute the string one-loop correction to the Einstein-Hilbert action. We write the graviton vertex operators
\begin{equation}
V_g^{(0)}=-\frac{2 g_c}{\alpha'}\epsilon_{\mu\nu} i\partial X^\mu \left(i\bar{\partial}X^\nu +\frac{\alpha'}{2}p\cdot \bar{\psi}\bar{\psi}^\nu\right) e^{i p\cdot X}\,,
\end{equation}
\begin{equation}
V_g^{(-1)}=-g_c \sqrt{\frac{2}{\alpha'}}\epsilon_{\mu\nu}\partial X^\mu e^{-\bar{\phi}}\bar{\psi}^\nu e^{ip\cdot X}\,,
\end{equation}
where $\epsilon_{\mu\nu}$ is the polarization tensor of the graviton. To compute the string-one loop correction to the Einstein-Hilbert term, we will compute a graviton two point function without imposing the momentum conservation condition. The reason for not imposing the momentum conservation condition is as follows. As in quantum field theory, S-matrix between two particle states should vanish on-shell because the two-point function is proportional to $p^2+m^2$ which vanishes when the momentum conservation condition is imposed.\footnote{In this paper, the on-shell condition always refers to $p^2+m^2=0.$} To read off the wave function renormalization of fields in QFT, one can factor out the factor that does not depend on the momentum. We shall illustrate this procedure using the graviton-graviton scattering amplitude. The graviton two-point function, before imposing the momentum conservation condition, is written as
\begin{equation}
Z=-AV_4 g_c^2 K(p_1,p_2,\epsilon_1,\epsilon_2)+\mathcal{O}(p^4)\,,
\end{equation}
where we define the kinematic factor
\begin{equation}\label{eqn:kin1}
\mathfrak{K}(p_1,p_2,\epsilon_1,\epsilon_2):= \left(p_{1\mu} p_{2\nu}\eta_{\rho\sigma}-p_{1\rho}p_{2\mu}\eta_{\nu\sigma} +\frac{1}{2}p_1\cdot p_2 \eta_{\mu\rho}\eta_{\nu\sigma} -\frac{1}{2}p_1\cdot p_2\eta_{\mu\nu}\eta_{\rho\sigma}\right)\epsilon^{\mu\nu}_1\epsilon_2^{\rho\sigma}\,.
\end{equation}
As one can check, the kinematic factor $\mathfrak{K}(p_1,p_2,\epsilon_1,\epsilon_2)$ vanishes if $p_1+p_2=0,$ because $p_1^2=p_2^2=0$ and $p_{1\mu}\epsilon_1^{\mu\nu}=p_{2\rho}\epsilon_2^{\rho\sigma}=0.$ Once $A$ is computed properly, one can read off the string one-loop correction to the Einstein-Hilbert term in string-frame \cite{Haack:2015pbv,Kim:2023sfs}
\begin{equation}
\frac{1}{2\kappa_4^2}\int d^4 x \left( \frac{\alpha'}{8\pi} A\right) R\,.
\end{equation}

Computing the full kinematic factor $\mathfrak{K}(p_1,p_2,\epsilon_1,\epsilon_2)$ is very cumbersome. To simplify the computation, one can impose the incomplete on-shell condition, which is equivalent to the transverse-traceless gauge condition
\begin{equation}\label{eqn:gauge 1}
p_1^2=p_2^2=p_1\cdot p_2=p_{1\mu}\epsilon_1^{\mu\nu}=p_{2\rho}\epsilon_2^{\rho\sigma} =\eta_{\mu\nu}\epsilon_1^{\mu\nu}=\eta_{\rho\sigma}\epsilon_2^{\rho\sigma}=0\,,
\end{equation}
while allowing $p_1+p_2=\xi$ to be a non-trivial null vector. After imposing this transverse-traceless gauge condition, we find 
\begin{equation}
\mathfrak{K}(p_1,p_2,\epsilon_1,\epsilon_2)=-p_{1\rho}p_{2\mu}\eta_{\nu\sigma} \epsilon_1^{\mu\nu}\epsilon_2^{\rho\sigma}\,,
\end{equation}
which is much simpler than \eqref{eqn:kin1}. We shall therefore impose \eqref{eqn:gauge 1} to compute $A.$ 

At the string-one loop, there are two distinct spin structures in the right-moving sector: even, and odd. We shall consider the even spin structures as the Einstein-Hilbert action is CP even. The torus amplitude is written as follows
\begin{align}
Z_{\mathcal{T}}=&\frac{V_4}{2^6\pi^4\alpha'^2}\sum_{\alpha,\beta \text{ even}} \int \frac{d^2\tau}{\tau_2^3}\int g_{z_1\bar{z}_1}d^2 z_1\int g_{z_2\bar{z}_2}d^2z_2\biggl[ \langle V_g^{(0,0)}(z_1,p_1,\epsilon_1) V_g^{(0,0)}(z_2,p_2,\epsilon_2)\rangle_\mathcal{T}^s\nonumber\\
&\qquad\qquad\qquad\qquad \times (-1)^{\alpha+\beta} \frac{\vartheta_{\alpha,\beta}(\bar{\tau})}{\eta(\tau)^2\eta(\bar{\tau})^3} \text{Tr}_\alpha\left((-1)^{\beta\bar{F}} q^{L_0-\frac{11}{12}}\bar{q}^{\bar{L}_0-\frac{3}{8}}\right)_{int}\biggr]\,,
\end{align}  
where the factor $1/(2^4\pi^4\tau_2^2)$ was introduced due to the integral over the momentum of the closed string
\begin{equation}
\int \frac{d^4k}{(2\pi)^4} e^{-\pi \tau_2\alpha'k^2}=\frac{1}{2^4\pi^4\alpha'^2 \tau_2^2}\,,
\end{equation}
and the factor $(-1)^{\alpha+\beta}\vartheta_{\alpha,\beta}(\bar{\tau})/(\eta(\tau)^2\eta(\bar{\tau})^3)$ comes from the non-compact directions and the ghost system, and we included an additional factor $1/2$ for the spin sum.

Now let us look at the two-point function more closely
\begin{equation}
-\frac{4g_c^2}{\alpha'^2}\epsilon^1_{\mu\nu}\epsilon^2_{\rho\sigma} \biggl\langle \partial X_1^\mu\left(i \bar{\partial}X^\nu_1+\frac{\alpha'}{2}p_1\cdot \bar{\psi}_1\bar{\psi}_1^\nu\right) \partial X_2^\rho \left(i \bar{\partial} X_2^\sigma +\frac{\alpha'}{2}p_2\cdot \bar{\psi}_2\bar{\psi}_2^\sigma\right)e^{ip_1\cdot X_1}e^{ip_2\cdot X_2} \biggr\rangle_\mathcal{T}^s\,.
\end{equation}
As we shall perform the spin sum for the right-moving sector, there should be contractions between fermions otherwise the contraction will vanish upon the spin sum. As a result, the only non-vanishing contribution comes from
\begin{equation}
-g_c^2\epsilon^1_{\mu\nu}\epsilon^2_{\rho\sigma} \biggl\langle \partial X_1^\mu  p_1\cdot \bar{\psi}_1\bar{\psi}_1^\nu \partial X_2^\rho p_2\cdot \bar{\psi}_2\bar{\psi}_2^\sigma e^{ip_1\cdot X_1}e^{ip_2\cdot X_2}\biggr\rangle_\mathcal{T}^s\,.
\end{equation}
Therefore, at order $\mathcal{O}(p_1\cdot p_2),$ the only non-vanishing contribution is from the following contraction
\begin{equation}
-g_c^2 \epsilon^1_{\mu\nu}\epsilon^2_{\rho\sigma} \langle \partial X_1^\mu \partial X_2^\rho \rangle_\mathcal{T}^s \,\langle p_1\cdot \bar{\psi}_1 \bar{\psi}_2^\sigma\rangle_\mathcal{T}^s\, \langle \bar{\psi}_1^\nu p_2\cdot \bar{\psi}_2\rangle_\mathcal{T}^s \,,
\end{equation}
which can be simplified to
\begin{equation}
-g_c^2 \epsilon^1_{\mu\nu}\epsilon^2_{\rho\sigma} \eta^{\mu\rho} p_1^\sigma p_2^\nu \langle \partial X_1\partial X_2\rangle_\mathcal{T} (\langle \bar{\psi}_1\bar{\psi}_2\rangle_\mathcal{T}^s)^2\,.
\end{equation}
To further simplify the expression, we shall use the following identities \cite{Alexandrov:2022mmy,Kim:2023sfs}
\begin{equation}
\left(\langle \bar{\psi}_1(0)\bar{\psi}_2(\bar{z})\rangle_\mathcal{T}^s\right)^2=\left(\frac{\vartheta_{\alpha,\beta}(\bar{z}|\bar{\tau})\vartheta_{1,1}'(0|\bar{\tau})}{\vartheta_{\alpha,\beta}(0|\bar{\tau})\vartheta_{1,1}(\bar{z}|\bar{\tau})}\right)^2\,,
\end{equation}
\begin{equation}\label{eqn:ferm 0}
\left(\frac{\vartheta_{\alpha,\beta}(\bar{z}|\bar{\tau})\vartheta_{1,1}'(0|\bar{\tau})}{\vartheta_{\alpha,\beta}(0|\bar{\tau})\vartheta_{1,1}(\bar{z}|\bar{\tau})}\right)^2=\frac{\vartheta_{\alpha,\beta}''(0|\bar{\tau})}{\vartheta_{\alpha,\beta}(0|\bar{\tau})}-\partial_{\bar{z}}^2 \log \vartheta_{1,1}(\bar{z}|\bar{\tau})\,,
\end{equation}
and
\begin{equation}\label{eqn:spin sum id0}
\sum_{\alpha,\beta}\vartheta_{\alpha,\beta}(0|\bar{\tau})\text{Tr}_\alpha\left((-1)^{\beta\bar{F}} q^{L_0-\frac{11}{12}}\bar{q}^{\bar{L}_0-\frac{3}{8}}\right)_{int}=0\,.
\end{equation}
Because of the identities \eqref{eqn:ferm 0} and \eqref{eqn:spin sum id0}, the vertex position-dependent terms in the fermion correlator drop out. Therefore, we can treat the fermion correlator as a constant concerning the vertex operator positions. Then, the vertex operator position integral reduces to
\begin{equation}\label{eqn:vertex integral}
\int g_{z_1\bar{z}_1}d^2z_2\int g_{z_2\bar{z}_2}d^2z_2 \langle \partial X_1(z_1)\partial X_2(z_2)\rangle\,.
\end{equation}
Note that the integrand in \eqref{eqn:vertex integral} is a total derivative! As a result, we conclude that the two-point function of the gravitons vanishes even off-shell $p_1+p_2\neq0.$ This implies that in heterotic string theory, the Einstein-Hilbert action is not renormalized at one-loop. This reproduces well known results in the literature \cite{Antoniadis:1992sa,Kiritsis:1994ta}.

Now, let us compute the dilaton two point function. The vertex operators for the dilaton are structurally similar to that of the graviton
\begin{equation}
V_D^{(0)}=-\frac{2g_c}{\alpha'} f_{\mu\nu} i\partial X^\mu \left(i\bar{\partial}X^\nu +\frac{\alpha'}{2}p\cdot \bar{\psi}\bar{\psi}^\nu\right)e^{ip\cdot X}\,,
\end{equation}
\begin{equation}
V_D^{(-1)}=ig_c \sqrt{\frac{2}{\alpha'}} f_{\mu\nu}\partial X^\mu e^{-\bar{\phi}}\bar{\psi}^\nu e^{ip\cdot X}\,,
\end{equation}
where the only difference lies in the polarization tensor
\begin{equation}
f_{\mu\nu}= \left(\eta_{\mu\nu} -\frac{1}{n\cdot p}(n_\mu p_\nu+n_\nu p_\mu)\right)\,,
\end{equation}
where $n$ is a generic vector such that $n\cdot p\neq0.$

To extract the one-loop correction to the dilaton kinetic term, we shall not impose the momentum conservation as in the case of the graviton-graviton scattering. But, in the case of the dilaton, we shall choose a different incomplete on-shell condition
\begin{equation}
p_1^2=p_2^2=f_{\mu\nu}^1 p_1^\mu=f_{\rho\sigma}^2p_2^\rho = 0\,,
\end{equation} 
and $p_1\cdot p_2\neq0.$ As our goal is then to compute the scattering amplitude of the form
\begin{equation}
-A V_4 g_c^2 p_1\cdot p_2+\mathcal{O}(p^3)\,,
\end{equation}
and read off the factor $A,$ without loss of generality we can impose a further condition $n_1=n_2=n.$ 

We shall focus on the correlation function $\langle V_D^{(0,0)}(z_1,p_1) V_D^{(0,0)}(z_2,p_2)\rangle_\mathcal{T}^s.$ The remainder of the computation is similar to the graviton scattering. We write the dilaton correlation function as
\begin{equation}
-\frac{4g_c^2}{\alpha'^2}f^1_{\mu\nu}f^2_{\rho\sigma} \biggl\langle \partial X_1^\mu\left(i \bar{\partial}X^\nu_1+\frac{\alpha'}{2}p_1\cdot \bar{\psi}_1\bar{\psi}_1^\nu\right) \partial X_2^\rho \left(i \bar{\partial} X_2^\sigma +\frac{\alpha'}{2}p_2\cdot \bar{\psi}_2\bar{\psi}_2^\sigma\right)e^{ip_1\cdot X_1}e^{ip_2\cdot X_2} \biggr\rangle_\mathcal{T}^s\,.
\end{equation}
Because of the spin sum over the even spin structures, we need at least four fermion contractions for the amplitude to be non-trivial. Therefore, we have
\begin{equation}\label{eqn:dilaton corr}
-g_c^2f^1_{\mu\nu}f^2_{\rho\sigma} \biggl\langle \partial X_1^\mu p_1\cdot \bar{\psi}_1\bar{\psi}_1^\nu \partial X_2^\rho p_2\cdot \bar{\psi}_2\bar{\psi}_2^\sigma e^{ip_1\cdot X_1}e^{ip_2\cdot X_2} \biggr\rangle_\mathcal{T}^s\,.
\end{equation}
We can further simplify \eqref{eqn:dilaton corr} into
\begin{equation}
-g_c^2 f_{\mu\nu}^1 f_{\rho\sigma}^2 \eta^{\mu\rho}(p_1^\sigma p_2^\nu -p_1\cdot p_2 \eta^{\nu\sigma})\langle \partial X_1\partial X_2\rangle_\mathcal{T} \left(\langle\bar{\psi}_1\bar{\psi}_2\rangle_\mathcal{T}^s\right)^2+\mathcal{O}(p^4)\,.
\end{equation}
As one can check
\begin{equation}
f_{\mu\nu}^1f_{\rho\sigma}^2 \eta^{\mu\rho}(p_1^\sigma p_2^\nu -p_1\cdot p_2 \eta^{\nu\sigma})=0\,,
\end{equation}
we conclude that the dilaton kinetic term is not renormalized at one-loop.

The computation we presented in this section implies that the kinetic term of the dilaton multiplet in Einstein frame is not renormalized at string one-loop. As we mentioned earlier, in the literature, it is well known that the one-loop correction to the K\"{a}hler potential depends on all moduli including the dilaton multiplet \cite{Antoniadis:1992rq,Antoniadis:1992pm,Kaplunovsky:1995jw}. Hence, it seems like we found a contradiction somehow. In the next section, we will show that there is no contradiction as the dilaton multiplet used in this section is a linear multiplet, whereas the commonly quoted one-loop correction to the K\"{a}hler potential is written in terms of chiral multiplets.
\subsection{Effective action in linear multiplet formalism}\label{sec:heterotic effective action}
In this section, we shall explain how the results found in the previous section can be reconciled with the well known results that in 4d $\mathcal{N}=1$ compactifications of heterotic string theory the K\"{a}hler potential is corrected at one-loop. A crucial insight that will help us resolve the puzzle is that the dilaton vertex operator used in the previous section corresponds to the variation of dilaton defined as a scalar component of a linear multiplet $L$ \cite{Derendinger:1994gx}. Note that the kinetic term of linear multiplets are not strictly speaking determined by the K\"{a}hler potential, although there is a similar function $K$ that determines the kinetic term. We will nevertheless abuse the langauge and call $K$ the K\"{a}hler potential in this section. We will decompose the tree level K\"{a}hler potential as
\begin{equation}
K=\mathcal{K}-\log (S+\bar{S})\,,
\end{equation}
or equivalently,
\begin{equation}
K=\mathcal{K}+\log (L)\,.
\end{equation}

A linear multiplet is defined by the following equations
\begin{equation}
(\mathcal{D}\mathcal{D} -8\mathcal{R}) L= (\bar{\mathcal{D}}\bar{\mathcal{D}}-8\mathcal{R})L=0\,,
\end{equation}
where $L$ is a real superfield, $\mathcal{D}$ and $\bar{\mathcal{D}}$ are covariant superderivatives defined as
\begin{equation}
\mathcal{D}_\alpha=\frac{\partial}{\partial \theta^\alpha}+i (\sigma^\mu \bar{\theta})_\alpha \partial_\mu\,,\quad \bar{\mathcal{D}}_{\dot{\alpha}}=-\frac{\partial}{\partial \bar{\theta}^{\dot{\alpha}}}-i(\theta\sigma^\mu)_{\dot{\alpha}}\partial_\mu\,.
\end{equation}
From the definition, it is clear where the name linear multiplet came from as the real supermultiplet $L$ is linear in the superspace coordinate when $\mathcal{R}=0.$ Supergravity theories that contain linear multiplets enjoy very rigid structures, due to the restrictions imposed on how to write supersymmetric actions in the presence of linear multiplets. Here, rather than providing an extensive overview of the linear multiplet formalism, we will instead summarize a few necessary ingredients to help us understand the K\"{a}hler potential.

The component form of the linear multiplet is given as \cite{Derendinger:1994gx}
\begin{equation}
L=l+i\theta \chi-i\bar{\theta}\bar{\chi}+\theta \sigma^\mu \bar{\theta}v_\mu-\frac{1}{2} \theta\theta\bar{\theta}(\partial_\mu\chi\sigma^\mu)-\frac{1}{2}\bar{\theta}\bar{\theta}\theta(\sigma^\mu \partial_\mu\bar{\chi})-\frac{1}{4}\theta\theta\bar{\theta}\bar{\theta} \partial^\mu\partial_\mu l\,,
\end{equation}
where $l$ is the scalar component, $\chi$ is the Majorana fermion, and $v_\mu$ is an axial vector field that satisfies
\begin{equation}
\partial_\mu v^\mu=0\,.
\end{equation}
Note that a divergent free vector field can be written as
\begin{equation}
v_\mu dx^\mu= dB\,,
\end{equation}
where $B$ is a two-form field. 

Now let us take a moment to digress into why this discussion of linear superfield is necessary. The bosonic components of the dilaton superfields are equivalent to the bosonic vertex operators obtained by varying the real rank two tensors of the worldsheet CFT. More precisely, the dilaton vertex operator is obtained as the variation of the trace of symmetric rank 2 tensor, whereas the B-field vertex operator is obtained as the variation of the anti-symmetric rank 2 tensor. As the derivation suggests, the dilaton vertex and the B-field vertex operator cannot be written as a real or imaginary part of a complex vertex operator. Hence, a combination of the dilaton vertex operator and the B-field vertex operator is necessarily a real vertex operator. As those two vertex operators form a supermultiplet, it is easy to understand why the dilaton multiplet that naturally arises in the string scattering amplitudes resides in the linear multiplet.

This observation that the dilaton multiplet, as seen from the string scattering amplitude computation, is a linear multiplet has interesting implications. Perhaps the most seemingly confusing implication is that the one-loop correction to the kinetic term of the dilaton seems to be absent, which defies our naive expectation that generically the kinetic term of every modulus is corrected at one-loop unless higher supersymmetry is providing a reason to believe that the kinetic term is protected. As we will see momentarily, this intuition relies on the assumption that all moduli fields are chiral multiplets, which are less restricted than the linear multiplet in a sense. 

We shall first study the effective action of linear multiplets, and later we will study how linear multiplets can be dualized to chiral multiplets. Once we understand the relation between the linear multiplet and the chiral multiplet, we will be able to understand the one-loop corrected K\"{a}hler potential. To simplify the discussion, let us first assume that the only low energy degrees of freedom are the gravity multiplet, chiral multiplets $S,$ and the linear multiplet $L.$ The two-derivative action of linear multiplets are determined by two functions $F$ and $K$ \cite{Binetruy:2000zx,Grimm:2005fa}
\begin{equation}\label{eqn:linear action0}
S=-\frac{3}{\kappa^2}\int E\, F(S,\bar{S},L)\,,
\end{equation}
where $E$ is the super-vielbein, the function $F$ depends on the K\"{a}hler potential $K(S,\bar{S},L)$ via the constraint
\begin{equation}\label{eqn:EH const}
1-\frac{L}{3} K_L=F-L F_L\,,
\end{equation}
which is imposed to obtain the correct normalization of the Einstein-Hilbert action in \emph{Einstein}-frame. If one wishes to go back to string-frame, one needs to modify the constraint \eqref{eqn:EH const}. The solution to \eqref{eqn:EH const} is
\begin{equation}
F(S,\bar{S},L)=1+L V(S,\bar{S})+\frac{L}{3}\int \frac{d\lambda}{\lambda} K_\lambda(S,\bar{S},\lambda)\,,
\end{equation}
where $V(S,\bar{S})$ can be thought of as an integration constant. Despite the name, $V(S,\bar{S})$ plays an important role, so we shall carefully treat this term. The component form of the action is then
\begin{equation}
S\supset -\frac{1}{2\kappa^2}\int d^4 x \left[ R-\frac{1}{2}\tilde{K}_{LL}\left( (\partial l)^2-(d B)^2\right)\right]\,.
\end{equation}
where we defined
\begin{equation}
\tilde{K}=K-3F\,.
\end{equation}

When a linear multiplet controls the coupling constant of a gauge multiplet $\mathcal{W},$ one should replace $L$ in the action \eqref{eqn:linear action0}
\begin{equation}
L\mapsto L+k_i \Omega^i\,,
\end{equation}
where $k_i$ is a constant that determines the overall normalization of the gauge coupling, and $\Omega^i$ is the Chern-Simons superfield defined by
\begin{equation}
\bar{D}^2\Omega^i=2 \text{tr}(\mathcal{W}^{i\alpha}\mathcal{W}^i_\alpha)\,\quad D^2\Omega=2\text{tr}(\mathcal{W}^i_{\dot{\alpha}}\mathcal{W}^{i\dot{\alpha}})\,.
\end{equation}
In the presence of the gauge fields, we have an additional term in the action given by
\begin{equation}\label{eqn:gauge}
-\frac{k}{4}\int d^4x \frac{F_L-\frac{1}{3} FK_L}{1-\frac{1}{3}L K_L} \left( \mathcal{D}^2 \text{tr}(\mathcal{W}^2)+h.c\right)\,,
\end{equation}
where the coupling constant is now then given by
\begin{equation}
-\frac{k}{4} \frac{F_L-\frac{1}{3}FK_L}{1-\frac{1}{3}LK_L}\,.
\end{equation}

To dualize the linear multiplet $L$ to a chiral multiplet $S_0,$ we can add a Lagrange multiplier to the action
\begin{equation}\label{eqn:dual action0}
S=-3\int E \,\left[ F(S,\bar{S},L)+(L-k_i \Omega^i)(S_0+\bar{S}_0)/6\right]\,.
\end{equation}
Varying the action \eqref{eqn:dual action0} with respect to $S_0$ and $\bar{S}_0$ gives a constraint equation that elliminates $S_0$ and $\bar{S}_0.$ On the other hand, if we vary \eqref{eqn:dual action0} with respect to $L,$ then we obtain a constraint equation
\begin{equation}\label{eqn:duality0}
\frac{S_0+\bar{S}_0}{6}\left(1-\frac{1}{3}LK_L\right)=\frac{1}{3} F K_L-F_L\,.
\end{equation}
Solving $L$ in terms of $S_0+\bar{S}_0,$ and plugging the solution back to \eqref{eqn:dual action0} then yields the supergravity action in terms of chiral multiplets $S_0$ and $S.$

We shall now go back to heterotic string compactification to understand the loop correction to the K\"{a}hler potential. Let us first study the effective action at the string-tree level. We will momentarily suppress the Chern-Simons form. At string tree-level, the K\"{a}hler potential in terms of $L,$ which is the 4d dilaton multiplet in heterotic compactification on a Calabi-Yau threefold $X_3,$ is given as
\begin{equation}
K(S,\bar{S},L)= \mathcal{K}(S,\bar{S})+\log L\,.
\end{equation}
Similarly, the prepotential $F$ is given as
\begin{equation}
F=\frac{2}{3}\,.
\end{equation}
The kinetic term of the linear multiplet is then given as
\begin{equation}
-\frac{1}{4\kappa^2}\int d^4 x \frac{1}{l^2} (\partial l)^2+\dots\,.  
\end{equation}
Now let us dualize the field $L$ into $S_0.$ We write the solution of \eqref{eqn:duality0} as
\begin{equation}
L= \frac{2}{(S_0+\bar{S}_0)}\,.
\end{equation}
In the dual formalism, the K\"{a}hler potential for moduli fields is now written as a more familiar form
\begin{equation}
K(S,\bar{S},S_0+\bar{S}_0)=\mathcal{K}(S,\bar{S})-\log(S_0+\bar{S}_0)\,,
\end{equation}
where the scalar component of $S_0$ is
\begin{equation}
s_0=e^{-2\phi} \mathcal{V}+i\int_{X_3} B_6\,,
\end{equation}
where $\mathcal{V}$ is a string-frame Calabi-Yau volume, and $\phi$ is the 10d dilaton. As one can see, the scalar component of $S_0$ scales as $g_s^{-2},$ whereas the scalar component of $L$ scales as $g_s^2.$

Now let us consider string one-loop correction to the effective action. As the string one-loop contribution must be suppressed by $g_s^2$ compared to the string tree-level contribution, we can parametrize the correction to the K\"{a}hler potential as
\begin{equation}
K(S,\bar{S},L)=\mathcal{K}(S,\bar{S})+\log L+ \mathcal{G}(S,\bar{S}) L\,,
\end{equation} 
and
\begin{equation}
F(S,\bar{S},L)=\frac{2}{3} +L\left(V(S,\bar{S}) +\frac{1}{3}\mathcal{G}(S,\bar{S})\log L\right)\,.
\end{equation}
Because the kinetic term of the linear multiplet is given by
\begin{equation}
\frac{1}{4\kappa^2}\int d^4x \tilde{K}_{LL}(\partial l)^2\,,
\end{equation}
the string one-loop corrected kinetic term is in general written as
\begin{equation}
-\frac{1}{4\kappa^2}\int d^4 x \frac{1+\mathcal{G}(S,\bar{S})l}{l^2} (\partial l)^2\,. 
\end{equation}
As we concluded in the previous section, $\mathcal{G}(S,\bar{S})$ vanishes identically, and the dilaton kinetic term is not renormalized at one-loop.

Does it mean that the kinetic term of the dilaton multiplet as a chiral multiplet is not renormalized at string one-loop? As we previously mentioned numerous times, the answer is no. We shall dualize the linear multiplet $L$ to $S_0,$ using the constraint equation
\begin{equation}
\frac{S_0+\bar{S}_0}{9}= \frac{1}{3L} \left( \frac{2}{3} +V(S,\bar{S})L\right) -V(S,\bar{S})\,.
\end{equation}
Therefore, the linear superfield $L$ is related to $S_0+\bar{S}_0$ as
\begin{equation}\label{eqn:sol duality0}
L=\frac{2}{S_0+\bar{S}_0+6 V(S,\bar{S})}\,.
\end{equation}
Plugging the solution \eqref{eqn:sol duality0} back to the action, we obtain that the K\"{a}hler potential for the chiral multiplet $S_0$ is simply
\begin{equation}
\mathcal{K}=\mathcal{K}(S,\bar{S})-\log (S_0+\bar{S}_0+6V(S,\bar{S}))\,,
\end{equation}
which indeed contains the string-loop correction 
\begin{equation}
\delta\mathcal{K}=-\frac{6V(S,\bar{S})}{S_0+\bar{S}_0}\,.
\end{equation}

We now observed that there is no contradiction between what we found in the previous section, and the well known result that there is indeed a non-vanishing string one-loop correction to the K\"{a}hler potential in 4d $\mathcal{N}=1$ compactifications of heterotic string theories. But, the remaining task is to compute $V(S,\bar{S})$ to complete the computation of string one-loop correction to the K\"{a}hler potential for chiral multiplets. The answer is extremely interesting. 

As was studied in \cite{Antoniadis:1992rq,Antoniadis:1992pm,Kiritsis:1994ta,Kaplunovsky:1995jw}, the threshold correction to the effective gauge coupling contains the information about $V(S,\bar{S})$
\begin{equation}
-\frac{k}{4}\int d^4 x \left(-\frac{1}{3L}+V(S,\bar{S})\right) \left( \mathcal{D}^2 \text{tr}(\mathcal{W}^2)+h.c\right)\,.
\end{equation}
By rewriting the action in the component forms, we find
\begin{equation}
-k\int d^4 x \left(-\frac{1}{3L}+V(S,\bar{S})\right) \text{tr} F^2\,.
\end{equation}
As a result, to determine the string one-loop correction to the K\"{a}hler potential, one can compute the universal part of the string threshold corrections to the gauge coupling! 

\subsection{Gauge threshold corrections}
To understand the universal part of the string threshold corrections, let us first compute string threshold corrections. Contents of this section are very well known in the literature \cite{Antoniadis:1992rq,Kaplunovsky:1992vs,Kiritsis:1994ta,Kaplunovsky:1995jw}. But, we decided to present them here for completeness. We write the gauge field vertex operator in the 0 picture as
\begin{equation}
V_A^{(0)}= \sqrt{\frac{2}{\alpha'}} \hat{k}^{-1/2} j^a e_\mu \cdot \left(i\bar{\partial}X^\mu +\frac{\alpha'}{2}  (p\cdot \psi)\psi^\mu\right) e^{ip\cdot X}\,.
\end{equation}
Although one can use the gauge field vertex operator, it is more convenient to use the vertex operator for the gauge field strength \cite{Kaplunovsky:1992vs}
\begin{equation}
V_F^{(0)}=\frac{i\alpha'}{4\pi} F_{\mu\nu} j^a \left(X^\mu\bar{\partial} X^\nu+\psi^\mu\psi^\nu\right)\,.
\end{equation}
To compute the gauge threshold correction, we compute the two-point function of the gauge field for the even spin structures
\begin{align}
Z_{\mathcal{T}}=&\frac{V_4}{2^6\pi^4\alpha'^2}\sum_{\alpha,\beta \text{ even}} \int \frac{d^2\tau}{\tau_2^3}\int g_{z_1\bar{z}_1}d^2 z_1\int g_{z_2\bar{z}_2}d^2z_2\biggl[ \langle V_F^{(0)}(z_1,p_1) V_F^{(0)}(z_2,p_2)\rangle_\mathcal{T}^s\nonumber\\
&\qquad\qquad\qquad\qquad \times (-1)^{\alpha+\beta} \frac{\vartheta_{\alpha,\beta}(\bar{\tau})}{\eta(\tau)^2\eta(\bar{\tau})^3} \text{Tr}_\alpha\left((-1)^{\beta\bar{F}} q^{L_0-\frac{11}{12}}\bar{q}^{\bar{L}_0-\frac{3}{8}}\right)_{int}\biggr]\,.
\end{align}  
As purely bosonic correlators vanish upon the spin sum, we can effectively treat the gauge field strength vertex operator as
\begin{equation}
\frac{i\alpha'}{4\pi} F_{\mu\nu} j^a \psi^\mu\psi^\nu\,.
\end{equation}
So we can simplify the two point function as
\begin{equation}
\langle V_F^{(0)}(z_1,p_1) V_F^{(0)}(z_2,p_2)\rangle_\mathcal{T}^s\supset \frac{\alpha'^2}{8\pi^2} F_{\mu\nu}^2 \langle j^a(z_1)j^a(z_2)\rangle_\mathcal{T}^s\left(\langle  \psi_1(\bar{z}_1)\psi_2(\bar{z}_2)\rangle_\mathcal{T}^s \right)^2\,.
\end{equation}
The resulting threshold correction is then written as
\begin{align}
\Delta=&-\frac{1}{2^6\pi^6}\sum_{\alpha,\beta \text{ even}} \int \frac{d^2\tau}{\tau_2^3}\int g_{z_1\bar{z}_1}d^2 z_1\int g_{z_2\bar{z}_2}d^2z_2\biggl[ \langle j^a(z_1)j^a(z_2)\rangle_\mathcal{T}^s\left(\langle  \psi_1(\bar{z}_1)\psi_2(\bar{z}_2)\rangle_\mathcal{T}^s \right)^2\nonumber\\
&\qquad\qquad\qquad\qquad \times (-1)^{\alpha+\beta} \frac{\vartheta_{\alpha,\beta}(\bar{\tau})}{\eta(\tau)^2\eta(\bar{\tau})^3} \text{Tr}_\alpha\left((-1)^{\beta\bar{F}} q^{L_0-\frac{11}{12}}\bar{q}^{\bar{L}_0-\frac{3}{8}}\right)_{int}\biggr]\,.
\end{align} 

We can further simplify the above result, first by using the identity \eqref{eqn:ferm 0} and
\begin{equation}
\langle j^a(z_1) j^a(z_2)\rangle = -k \partial_z^2 \log\vartheta_{1,1}(z_1-z_2|\tau)+\pi^2(Q^a)^2\,,
\end{equation}
we obtain
\begin{equation}\label{eqn:threshold1}
\Delta=-\frac{1}{2^6\pi^4}\sum_{\alpha,\beta\text{ even}}\int\frac{d^2\tau}{\tau_2} (-1)^{\alpha+\beta}\frac{\vartheta_{\alpha,\beta}''(\bar{\tau})}{\eta(\tau)^2\eta(\bar{\tau})^3}\left(Q^2-\frac{\pi k}{2\tau_2}\right)
\text{Tr}_\alpha\left((-1)^{\beta\bar{F}} q^{L_0-\frac{11}{12}}\bar{q}^{\bar{L}_0-\frac{3}{8}}\right)_{int}\,.
\end{equation}
Using the heat kernel equation
\begin{equation}
\vartheta_{1,1}''(z|\tau)=4\pi i\frac{\partial}{\partial \tau}\vartheta_{1,1}(z|\tau)\,,
\end{equation}
we can rewrite \eqref{eqn:threshold1} in a more well known form \cite{Kaplunovsky:1992vs}
\begin{equation}\label{eqn:threshold2}
\Delta_a=\frac{1}{2^4\pi^2}\sum_{\alpha,\beta\text{ even}}\int\frac{d^2\tau}{\tau_2} (-1)^{\alpha+\beta} \frac{dZ_{\psi}(s,\bar{\tau})}{\pi i|\eta(\tau)|^4d\bar{\tau}}\left(Q_{(a)}^2-\frac{\pi k}{2\tau_2}\right)
\text{Tr}_\alpha\left((-1)^{\beta\bar{F}} q^{L_0-\frac{11}{12}}\bar{q}^{\bar{L}_0-\frac{3}{8}}\right)_{int}\,,
\end{equation}
or \cite{Antoniadis:1992rq,Antoniadis:1992pm}
\begin{equation}\label{eqn:threshold3}
\Delta_a=\frac{1}{2^4\pi^2}\int\frac{d^2\tau}{\tau_2} (-1)^{\alpha+\beta}\frac{1}{\eta(\tau)^2} \left(Q_{(a)}^2-\frac{\pi k}{2\tau_2}\right)
\text{Tr}_{\bar{R}}\left((-1)^{\bar{F}} \bar{F}q^{L_0-\frac{11}{12}}\bar{q}^{\bar{L}_0-\frac{3}{8}}\right)_{int}\,.
\end{equation}
Note that in the context of the 4d $\mathcal{N}=2$ compactification, the threshold correction was shown to be determined by the BPS data of the K3 CFT \cite{Harvey:1995fq}.

We have now computed the gauge group-dependent threshold corrections. As was studied in great detail in a seminal paper \cite{Kaplunovsky:1995jw}, the Green-Schwarz term is equivalent to the universal threshold correction. The derivation of the equivalence between the GS term and the universal threshold correction is quite involved. So, rather than replicating the derivation, we will sketch the crucial ideas used in the derivation.

The threshold correction we computed in \eqref{eqn:threshold1} does not necessarily have to be equal to the threshold correction that one can compute within low-energy supergravity. For gauge group $G_a,$ we denote the threshold correction that is completely fixed within the low energy supergravity by
\begin{equation}
\tilde{\Delta}_a\,.
\end{equation}
The gauge coupling measured at the string-scale is
\begin{equation}
\frac{1}{g_{string}^{2}}=\re S+\frac{1}{16\pi^2}\Delta^{univ}\,,
\end{equation}
which determined the relation between $\Delta_a,$ $\tilde{\Delta}_a,$ and $\Delta^{univ}$
\begin{equation}
\tilde{\Delta}_a=\Delta_a+k_a\Delta^{univ}\,,
\end{equation}
where $k_a$ is the level of the Kac-Moody current algebra. 

As was studied in \cite{Kaplunovsky:1995jw}, $\tilde{\Delta}_a$ is related to the K\"{a}hler potential through the Kaplunovsky-Louis (KL) formula \cite{Kaplunovsky:1994fg}. We shall review how the relationship was obtained in \cite{Kaplunovsky:1995jw}. Let us start with the KL formula
\begin{equation}
\frac{1}{g_a^2(\mu)}=k_a \re S +\frac{b_a}{16\pi^2} \log\frac{M_{pl}^2}{\mu^2}+\frac{c_a}{16\pi^2} \mathcal{K} +\frac{T(G_a)}{8\pi^2}\log g_a^{-2}(\mu)-\sum_r\frac{T_a(r)}{8\pi^2}\log\det Z_{(r)}(\mu^2)\,,
\end{equation}
where $\mathcal{K}$ is the tree-level moduli K\"{a}hler potential, $Z_{(r)}$ is the tree-level matter kinetic term in an irreducible representation $r,$ 
\begin{equation}
c_a=\sum T_a(r)-T(G_a)\,,
\end{equation}
and
\begin{equation}
b_a=\sum T_a(r)-3T(G)\,.
\end{equation}
We can relate the Planck mass to the string mass scale by using
\begin{equation}
M_{pl}^2=M_{st}^2 \re S\,,
\end{equation}
to rewrite the KL formula as
\begin{equation}\label{eqn:KL het}
\frac{1}{g_a^2(\mu)}=k_a\re S+\frac{b_a}{16\pi^2}\log \frac{M_{st}^2}{\mu^2}+\frac{\re f_a^{(1)}}{16\pi^2}+\frac{c_a}{16\pi^2} \hat{\mathcal{K}} -\sum_r\frac{T_a(r)}{8\pi^2}\log\det Z_{(r)}^{(tree)}\,,
\end{equation}
where $\hat{\mathcal{K}}$ is defined as
\begin{equation}
\hat{\mathcal{K}}:=\mathcal{K}+\log(S+\bar{S})\,.
\end{equation}

The field theoretical threshold correction then can be obtained by identifying the field theoretical KL formula with the stringy threshold computation
\begin{equation}\label{eqn:KL stringy het}
\frac{1}{g_a^2(\mu)}= \frac{1}{g_a^{2}(M_{st})}+\frac{b_a}{16\pi^2}\log \frac{M_{st}^2}{\mu^2}+\frac{\Delta_a}{16\pi^2}\,,
\end{equation}
where the gauge coupling at string scale at one-loop is defined as
\begin{equation}
\frac{1}{g_a^2(M_{st})}=k_a\re S+\frac{k_a}{16\pi^2}\Delta^{univ}\,.
\end{equation}
We identify \eqref{eqn:KL stringy het} with \eqref{eqn:KL het} to obtain the formula we were after
\begin{equation}
\partial_{\phi}\bar{\partial}_{\bar{\phi}}\tilde{\Delta}_a=\partial_{\phi}\bar{\partial}_{\bar{\phi}}\left( c_a \mathcal{K}-\sum_r2\text{Tr}_a(r)\log \det Z_{(r)}^{tree}\right)\,.
\end{equation}

We are not quite done yet, because we need to understand how $\tilde{\Delta}_a$ is related to the Green-Schwarz term. To relate $\tilde{\Delta}_a$ to the Green-Schwarz term that parametrizes the string one-loop correction to the K\"{a}hler potential, one can compute a scattering amplitude between two gauge bosons and one moduli field $M.$ In the low energy supergravity computation, this amplitude is proportional to
\begin{equation}\label{eqn:sugra 2A1M}
\pm\frac{i\epsilon_{\mu\alpha\nu\beta}p_1^\alpha p_2^\beta}{16\pi^2}\times \frac{\partial}{\partial M} \left( \tilde{\Delta}_a -48\pi^2 k_a V\right)\,.
\end{equation}
Note that the Green-Schwarz term enters the scattering amplitude because the Green-Schwarz term mediates the kinetic mixing between $S$ and $M.$ On the other hand, the stringy computation of the same scattering amplitude is given as \cite{Antoniadis:1992rq}
\begin{equation}\label{eqn:stringy 2A1M}
\pm\frac{i\epsilon_{\mu\alpha\nu\beta}p_1^\alpha p_2^\beta}{16\pi^2}\times \frac{\partial}{\partial M} \Delta_a\,.
\end{equation}
By equating \eqref{eqn:sugra 2A1M} with \eqref{eqn:stringy 2A1M}, we arrive at the relation
\begin{equation}
\Delta^{univ}-48\pi^2V=0\,.
\end{equation}
As a result, one can translate the stringy one-loop threshold correction to the Green-Schwarz term
\begin{equation}
V=\frac{1}{48\pi^2 k_a} \left[-\Delta_a+c_a\mathcal{K}-\sum_r2\text{Tr}_a(r)\log\det Z_{(r)}^{tree}\right]+\text{harmonic terms}\,.
\end{equation}
The relation between the Green-Schwarz term and the Kahler potential in type II string theory will be studied elsewhere \cite{GS}.

\section{Type II string theories}\label{sec:type II}
In this section, we shall study the string one-loop corrections to the kinetic action of the 4d dilaton in type II compactifications on Calabi-Yau orientifolds. Throughout most of the section, we will keep the discussion quite general.

We shall begin by describing the worldsheet CFT. To obtain $\mathcal{N}=1$ spacetime supersymmetry, we shall perform orientifolding. To find consistent orientifolds, one needs to saturate the RR-tadpole. In this work, we shall saturate the RR-tadpole with spacetime-filling D-branes. For the details of the orientifolding, see \cite{Alexandrov:2022mmy}. We will mostly follow the orientifold conventions of \cite{Alexandrov:2022mmy}. In general, D-branes and O-planes in the spectrum will backreact on, and the backreaction will cause the non-trivial Ramond-Ramond profile even if the RR-tadpole is canceled. To make progress, we shall work in the regime in which such backreactions are small, which can be understood as the small warping limit \cite{Kim:2023sfs}. 

In the small warping limit, we have an exact description of the worldsheet CFT. The worldsheet CFT consists of the ghost CFT and the matter CFT. The ghost CFT contains the usual $b,~c,~\beta,~\gamma$ ghost system. The matter part of the worldsheet theory contains $(1,1)$ supersymmetric free field CFT with the central charge $(c,\bar{c})=(6,6)$ for the non-compact directions, and a $(2,2)$ supersymmetric CFT with the central charge $(c,\bar{c})=(9,9)$ for the Calabi-Yau compactification. We will use $X^\mu$ and $\psi^\mu$ to denote the usual bosonic and fermionic free fields in the matter CFT. D-branes and O-planes will be described as the boundary states of the aformentioned CFT.

Because the kinetic term of the 4d dilaton is CP-even, we only need to sum over the even spin structures for open strings, and the (even,even) and (odd,odd) spin structures for closed strings. Before writing the dilaton vertex operator, let us first define auxiliary vertex operators 
\begin{equation}
V^{(0)\mu}(z,p)=\sqrt{\frac{2}{\alpha'}}\left(i\partial X(z)+\frac{\alpha'}{2}p\cdot \psi \psi^\mu (z)\right) e^{i p\cdot X(z)}\,,
\end{equation}
and
\begin{equation}
V^{(-1)\mu}(z,p)=- e^{-\phi(z)}\psi^\mu(z) e^{ip\cdot X(z)}\,.
\end{equation}
It is important to note that the vertex operators $V^{(0)\mu}(z,p)$ and $V^{(-1)\mu}(z,p)$ are holomorphic. Similarly, we define the anti-holomorphic version of the auxiliary vertex operators
\begin{equation}
\overline{V}^{(0)\mu}(\bar{z},p)=\sqrt{\frac{2}{\alpha'}}\left(i\bar{\partial} X(\bar{z})+\frac{\alpha'}{2}p\cdot \bar{\psi} \bar{\psi}^\mu (\bar{z})\right) e^{i p\cdot \tilde{X}(\bar{z})}\,,
\end{equation}
and
\begin{equation}
\overline{V}^{(-1)\mu}(\bar{z},p)=- e^{-\bar{\phi}(\bar{z})}\bar{\psi}^\mu(\bar{z}) e^{ip\cdot \tilde{X}(\bar{z})}\,.
\end{equation}
We write the dilaton vertex operators in the $(p,q)$ picture as \cite{Klebanov:1995ni,Garousi:1996ad,Hashimoto:1996bf}
\begin{equation}
V_D^{(p,q)}(z,\bar{z},p)=-g_c f_{\mu\nu} :V^{(p)\mu}(z,p): :\overline{V}^{(q)\nu}(\bar{z},p):\,,
\end{equation}
where the polarization tensor is given as
\begin{equation}
f_{\mu\nu}= \left(\eta_{\mu\nu}-\frac{1}{n\cdot p}(n_\mu p_\nu+p_\mu n_\nu)\right)\,,
\end{equation}
and $n$ is a generic four vector for which $n\cdot p\neq 0.$ We shall choose the auxiliary vector $n$ such that $n^2=0$ \cite{Blumenhagen:2013fgp}. The integrated vertex operator is then written as
\begin{equation}
\int d^2 z V_D^{(p,q)}(z,\bar{z},p)\,.
\end{equation}

To check the normalization of the dilaton vertex operator, we can use the normalization of the dilaton vertex operator which was computed using a disk diagram ending on a D(-1)-brane \cite{Alexandrov:2021dyl}. Properly normalized dilaton vertex operator for $\phi_4$ is 
\begin{equation}
V^{(-1,-1)}_{\phi_4}=\frac{1}{\kappa_4}V_{D}^{(-1,-1)}\,,
\end{equation}
after taking into account the difference in conventions between this draft and \cite{Alexandrov:2021dyl}. As a result we conclude that $V_D$ is the vertex operator for $\kappa_4 \phi_4.$ One can arrive at the same conclusion by using the following hack. To obtain the canonically normalized vertex operator in Einstein-frame at string tree-level, one can vary $\phi_4=\phi_4^*+\kappa_4 \delta \phi_4 e^{ip\cdot x}/\sqrt{2}$ in Einstein-frame. Translating this into string-frame, one obtains the variations $g_{\mu\nu}=\eta_{\mu\nu}+2\kappa_4 \eta_{\mu\nu} e^{ip\cdot x}/\sqrt{2}$ and $\phi_4=\phi_4^{*}+\kappa_4 \delta \phi_4 e^{ip\cdot x}/\sqrt{2}.$ This corresponds to the covariant dilaton vertex operator with the polarization tensor $\epsilon_{\mu\nu}=\eta_{\mu\nu}/\sqrt{2}$ \cite{Polchinski:1998rq}. By going to the light-cone gauge, we therefore find that the canonically normalized field corresponds to $\epsilon_{\mu\nu}=(\eta_{\mu\nu}-(n_\mu p_\nu-n_\nu p_\mu)/(n\cdot p))/\sqrt{2}.$ Therefore, we again conclude that the vertex operator $V_D$ corresponds to the variation $\kappa_4\delta\phi_4.$

Note that we will oftentimes use a shorthand notation
\begin{equation}
X(z,\bar{z})=X(z)+\tilde{X}(\bar{z})\,.
\end{equation}
In order to simplify the expressions, from now on, we will omit the normal-ordering symbols. But, keeping track of the correct normal ordering will be very important when we are computing the self-contraction terms.\footnote{For example, in a recent study of D-instanton amplitudes in type IIB string theory, careful treatment of such self-contractions was absolutely crucial to obtain the correct answer \cite{Agmon:2022vdj}.} For the worldsheet conventions used in this paper, see \S\ref{sec:Green function}.

In the next sections, we shall compute the dilaton two point functions.

\subsection{Even spin structures}
In this section, we will study the (even,even) spin structures for closed string worldsheets and the even spin structures for open string worldsheets. As most of the discussion is analogous, we will not specify which type of worldsheet is considered and keep the computation very general. Note that the Chan-Paton factors are included in the internal part of the CFT.

We shall compute the two point function
\begin{equation}
Z_\sigma(p_1,p_2)=\langle\langle V_D^{(0,0)}(z_1,\bar{z}_1,p_1)V_D^{(0,0)}(z_2,\bar{z}_2,p_2)\rangle\rangle_\sigma^s\,,
\end{equation}
for Riemann surfaces $\sigma$ of genus 1 with the even spin structures $s.$ In this section, we will only need explicit expressions for the partition function for $\sigma\neq \mathcal{T}.$ The reason is quite simple, because the torus amplitude with the even spin structure simply vanishes. This point will become very clear later in this section. We write
\begin{align}
Z_\sigma=& \frac{c_\sigma V_4}{ 2^6\pi^4\alpha'^2}\sum_{\alpha,\beta \text{ even}}\int_0^\infty \frac{dt}{t^3}\int_\sigma  d^2z_1\int_\sigma  d^2z_2 \bigg[ \langle V_D^{(0,0)}(z_1,\bar{z}_1,p_1)V_D^{(0,0)}(z_2,\bar{z}_2,p_2)\rangle_\sigma^s\nonumber\\
&\qquad\qquad\qquad\qquad\qquad\times(-1)^{\alpha+\beta} \frac{\vartheta_{\alpha,\beta}(\tau)}{\eta(\tau)^3} \text{Tr}_\alpha \left((-1)^{\beta F}q^{L_0-\frac{3}{8}} \right)^{\sigma}_{int}\biggr]\,,
\end{align}
where $c_\sigma=2^{-2}$ for annulus and M\"{o}bius strip, and $c_\sigma=1$ for Klein bottle. The factor $c_\sigma/(2^4\pi^4t^2)$ was included due to the integral over the momentum of the string, and the factor $(-1)^{\alpha+\beta}\vartheta_{\alpha,\beta}(\tau)/\eta(\tau)^3$ comes from the non-compact directions and the ghost system, and we included an additional factor of $1/2$ for the GSO projection. As was emphasized, we will not impose the on-shell condition $p_1\cdot p_2=0.$ But, we will make sure that the states $p_1$ and $p_2$ are on-shell $p_1^2=p_2^2=0.$ 

We write the dilaton correlator as
\begin{align}
\mathfrak{C}_\sigma:=&\langle V_D^{(0,0)}(z_1,\bar{z}_1,p_1)V_D^{(0,0)}(z_2,\bar{z}_2,p_2)\rangle_\sigma^s\,,\\
=& \frac{4g_c^2}{\alpha'^2}f_{\mu\nu}^1f_{\rho\sigma}^2\left\langle e^{ip_1\cdot X_1}e^{ip_2\cdot X_2} \left(i\partial X_1^\mu +\frac{\alpha'}{2}p_1\cdot \psi_1\psi_1^\mu\right)\left(i\bar{\partial}X_1^\nu+\frac{\alpha'}{2}p_1\cdot\bar{\psi}_1\bar{\psi}_1^\nu\right) \right.\nonumber\\
&\qquad\qquad\qquad\times\left. \left(i\partial X_2^\rho+\frac{\alpha'}{2}p_2\cdot \psi_2 \psi_2^\rho\right)\left(i\bar{\partial}X_2^\sigma+\frac{\alpha'}{2}p_2\cdot \bar{\psi}_2\bar{\psi}_2^\sigma\right)  \right\rangle_\sigma^s\,.
\end{align}
A few comments are in order. As was emphasized, we aim to extract a term that is proportional to $\delta:=\alpha'p_1\cdot p_2$ in the small $\delta$ limit. Bosonic correlators do not depend on the spin structure. As a result, purely bosonic contractions will vanish after summing over the spin structures. This can be understood as a result of the target spacetime supersymmetry. Therefore, there should be at least four fermion contractions. Naively, one can think that the eight fermion contraction will not contribute to $\mathcal{O}(\delta),$ because contracting eight fermions will produce $\delta^2$ factors. But, it is possible that the integral over moduli will produce a $1/\delta$ factor due to vertex collisions. In fact, in the context of toroidal orientifold compactifications, it was observed that such vertex collisions can occur \cite{Berg:2014ama}. Therefore, by adopting the technology developed in \cite{Berg:2014ama}, we shall study the eight-fermion contributions carefully.

We close this section with one more comment. One can also worry that the moduli integral in the UV region $t\rightarrow 0$ can provide an additional source of $1/\delta$ enhancement as in \cite{Antoniadis:2002cs}.\footnote{Note that in \cite{Antoniadis:2002cs}, only the annuli diagrams were considered. But, in our paper, we sum over all worldsheet topologies of genus 1. This will play an important role.} To examine such an enhancement, let us go to the closed string channel by changing the worldsheet modulus to $l:=1/t$ for annuli,  $l:=1/(4t)$ for the Mobius strip, and $l:=1/(2t)$ for Klein bottle \cite{Polchinski:1998rq,Polchinski:1998rr}. Then, the one-loop diagram in the UV region, $t\rightarrow 0,$ can receive contributions of the form
\begin{equation}
\delta^2\int_1^{\infty} dl \biggl[l^{a+\delta} e^{-\pi (L_0+\bar{L_0}) l}\biggr]\,,
\end{equation}
where $a$ can be either 1 or 0 in the amplitudes considered in this paper. Then, upon the integral, one finds that
\begin{equation}
\delta^2\int_1^{\infty} dl \biggl[l^{a+\delta} e^{-\pi (L_0+\bar{L_0}) l}\biggr]=\frac{\delta^2}{(L_0+\bar{L}_0)^{1+a+\delta}}\Gamma(1+a)+\dots\,.
\end{equation}
If there are massless states in, say, the annulus, then the massless state exchange will produce 
\begin{equation}
L_0+\bar{L}_0= \delta\,.
\end{equation}
Which, therefore, can produce order $\mathcal{O}(\delta)$ contribution for $a=0$
\begin{equation}
\delta^2\int_1^{\infty} dl \biggl[l^{\delta} e^{-\pi (L_0+\bar{L_0}) l}\biggr]=\delta\times\Gamma(1)+\dots\,.
\end{equation}
Such an enhancement can indeed occur in an \emph{individual} diagram. But, it is never enough to emphasize that in a consistent theory where the tadpole is cancelled, sum over the worldsheet topologies will project out the massless states in the closed string channel of the one-loop amplitudes by definition. Otherwise, in the one-loop partition function of string theory, there will be a contribution of the form
\begin{equation}
\int^\infty_1 dl \left[ N_{massless}+\mathcal{O}\left(e^{-2l}\right)\right]\,,
\end{equation}
which will force the one-loop partition function to diverge. As a result, we conclude that we don't expect to see the enhancment by a factor of $1/\delta$ from the moduli integral around the UV region $t\rightarrow 0.$

\subsubsection{Four fermion terms}
Let us first study the four-fermion terms. We shall first distinguish the ``self-contractions'' from genuine contractions between fields in the two different vertex operators
\begin{equation}
V_D^{(0,0)}(z_1,\bar{z}_1,p_1)\,,
\end{equation}
and 
\begin{equation}
V_D^{(0,0)}(z_2,\bar{z}_2,p_2)\,.
\end{equation}
The self-contraction here refers to a contraction between fields within a single vertex operator. For example, there can be contractions $\langle p_1\cdot \psi_1 p_1\cdot \bar{\psi}_1\rangle$ or $\langle p_1\cdot \psi_1 \bar{\psi}_1^\nu\rangle.$ Because the dilaton vertex operator is normal ordered, not boundary normal ordered, in general such self contractions can be non-vanishing. But, it is important to note that in fact such self contractions at four fermion level will generate zero contributions, because we are imposing the mass-shell conditions
\begin{equation}
p_1^2=p_2^2=0\,,
\end{equation}
and the transversality conditions
\begin{equation}
f_{\mu\nu}^1p_1^\mu =f_{\mu\nu}^2p_2^\mu=0\,.
\end{equation}
In the case of the eight-fermion contractions, we will see that the self-contractions will produce interesting results.

Focusing on the genuine contractions, there are in total four different types of contractions. Because all of them have very similar structures, we will focus on the contribution due to a contraction involving $\partial X_1^\mu $ and $\partial X_2^\rho,$ and record the other contributions. We compute
\begin{equation}
\mathfrak{C}_4^{(1)}:=-g_c^2f_{\mu\nu}^1f_{\rho\sigma}^2\langle \partial X_1^\mu\partial X_2^\rho e^{ip_1\cdot X_1}e^{i p_2\cdot X_2}\rangle \langle p_1\cdot\bar{\psi}_1\bar{\psi}_1^\nu p_2\cdot\bar{\psi}_2\bar{\psi}_2^\sigma\rangle\,.
\end{equation}
There are two different types of contractions among fermions
\begin{equation}\label{eqn:four ferm c1}
\mathfrak{C}_4^{(1)}=-g_c^2f_{\mu\nu}^1f_{\rho\sigma}^2\langle\partial X_1^\mu\partial X_2^\rho e^{ip_1\cdot X_1}e^{i p_2\cdot X_2}\rangle \left(\langle p_1\cdot \bar{\psi}_1 \bar{\psi}_2^\sigma \rangle \langle \bar{\psi}_1^\nu p_2\cdot \bar{\psi}_2 \rangle- \langle p_1\cdot \bar{\psi}_1p_2\cdot \bar{\psi}_2\rangle \langle \bar{\psi}_1^\nu\bar{\psi}_2^\sigma\right)\,.
\end{equation}

Naively, one might think that we can simply drop the exponential factors
\begin{equation}
e^{ip_1\cdot X_1}e^{ip_2\cdot X_2}\,,
\end{equation}
and ignore the contractions between $\partial X$ and the exponential factors because such contractions will produce higher order terms in the momentum. But, as we previously mentioned, moduli integral around a pole induced by vertex collisions can bring down a factor of $1/\delta.$ So, we should also pay extra attention to the contractions between $\partial X$ and the exponential factors. 

Let us first focus on the contractions between $\partial X.$ We will later return to the contractions between the exponential factors and $\partial X.$ Using the relations
\begin{equation}
\langle \partial X_1^\mu \partial X_2^\nu\rangle = \eta^{\mu\nu}\langle \partial X_1 \partial X_2\rangle\,,\quad \langle \bar{\psi}_1^\mu\bar{\psi}_2^\nu\rangle = \eta^{\mu\nu}\langle\bar{\psi}_1\bar{\psi}_2\rangle\,,
\end{equation}
and ignoring the exponential factors, we obtain
\begin{equation}
\mathfrak{C}_{4,n}^{(1)}=-g_c^2f_{\mu\nu}^1f_{\rho\sigma}^2 \eta^{\mu\rho} ( p_1^\sigma p_2^\nu- p_1\cdot p_2 \eta^{\nu\sigma})\langle \partial X_1\partial X_2\rangle \langle \bar{\psi}_1\bar{\psi}_2\rangle^2\,.
\end{equation}

We first compute 
\begin{align}
\mathfrak{K}_1:=& f_{\mu\nu}^1 f_{\rho\sigma}^2\eta^{\mu\rho} p_1^\sigma p_2^\nu\,,\\
=&\left(\eta_{\mu\nu}-\frac{1}{n\cdot p_1} (n_\mu p_{1\nu}+n_{\nu}p_{1\mu})\right)\left(\eta_{\rho\sigma}-\frac{1}{n\cdot p_2}(n_\rho p_{2\sigma}+n_\sigma p_{2\rho})\right)\eta^{\mu\rho} p_1^\sigma p_2^\nu\,,\\
=&2 p_1\cdot p_2\,. 
\end{align}
Similarly, we compute
\begin{align}
\mathfrak{K}_2:=& p_1\cdot p_2f_{\mu\nu}^1f_{\rho\sigma}^2\eta^{\mu\rho} \eta^{\nu\sigma}\,,\\
=&p_1\cdot p_2\left(\eta_{\mu\nu}-\frac{1}{n\cdot p_1} (n_\mu p_{1\nu}+n_{\nu}p_{1\mu})\right)\left(\eta_{\rho\sigma}-\frac{1}{n\cdot p_2}(n_\rho p_{2\sigma}+n_\sigma p_{2\rho})\right) \eta^{\mu\rho}\eta^{\nu\sigma}\,,\\
=&2p_1\cdot p_2\,.
\end{align}
As a result, for the kinematic reason we conclude
\begin{equation}
\mathfrak{C}_4^{(1)}=0\,.
\end{equation}
Because the polarization tensors $\epsilon^1$ and $\epsilon^2$ are symmetric, we further conclude that all combinations vanish. Hence, the four-fermion contributions cannot yield a non-trivial result from the contractions between non-exponentiated bosonic fields.

Now we shall study the contractions between $\partial X$ and $e^{ip\cdot X}$ factors. We will find that unlike the non-universal moduli and graviton amplitudes \cite{Berg:2005ja,Berg:2014ama,Haack:2015pbv,Kim:2023sfs}, the dilaton amplitude is very special in that it will receive non-trivial corrections due to the fact that the vertex collision can produce the enhancement by the factor of $1/\delta$ even at the four fermion level. Because of the transversality condition of the polarization, in \eqref{eqn:four ferm c1}, $\partial X_1^\mu$ can only be contracted with $e^{ip_2\cdot X_2}$ and similarly $\partial X_2^\rho$ can only be contracted with $e^{ip_1\cdot X_1}.$ By using the relation
\begin{equation}
\langle \partial X_1^\mu \partial X_2^\rho e^{ip_1\cdot X_1}e^{ip_2\cdot X_2}\rangle=  \left(\eta^{\mu\rho} \langle \partial X_1\partial X_2\rangle -p_1^\rho p_2^\mu \langle \partial X_1 X_2\rangle \langle X_1\partial X_2\rangle \right) e^{-p_1\cdot p_2\langle X_1X_2\rangle}\,,
\end{equation}
we obtain that the contraction between $\partial X$ and the exponential factor yields
\begin{equation}\label{eqn:four ferm c1 2}
\mathfrak{C}_{4,e}^{(1)}=g_c^2f_{\mu\nu}^1f_{\rho\sigma}^2 p_1^\rho p_2^\mu (p_1^\sigma p_2^\nu-p_1\cdot p_2 \eta^{\nu\sigma}) \langle \partial X_1 X_2\rangle \langle X_1\partial X_2\rangle \langle \bar{\psi}_1\bar{\psi}_2\rangle^2e^{-p_1\cdot p_2\langle X_1X_2\rangle}\,.
\end{equation}
Finally, we use the following identity 
\begin{equation}
f_{\mu\nu}^1f_{\rho\sigma}^2p_1^\rho p_2^\mu(p_1^\sigma p_2^\nu-p_1\cdot p_2 \eta^{\nu\sigma})=2(p_1\cdot p_2)^2\,,
\end{equation}
to simplify \eqref{eqn:four ferm c1 2} into
\begin{equation}\label{eqn:four ferm c1 0}
\mathfrak{C}_{4,e}^{(1)}=2g_c^2(p_1\cdot p_2)^2 \langle \partial X_1 X_2\rangle \langle X_1\partial X_2\rangle \langle \bar{\psi}_1\bar{\psi}_2\rangle^2e^{-p_1\cdot p_2\langle X_1X_2\rangle}\,.
\end{equation}

Similarly, from
\begin{equation}
\mathfrak{C}_4^{(2)}:=-g_c^2f_{\mu\nu}^1f_{\rho\sigma}^2\langle \partial X_1^\mu\bar{\partial} X_2^\sigma e^{ip_1\cdot X_1}e^{i p_2\cdot X_2}\rangle \langle p_1\cdot\bar{\psi}_1\bar{\psi}_1^\nu p_2\cdot\psi_2\psi_2^\rho\rangle\,,
\end{equation}
\begin{equation}
\mathfrak{C}_4^{(3)}:=-g_c^2f_{\mu\nu}^1f_{\rho\sigma}^2\langle \partial \bar{X}_1^\nu\partial X_2^\rho e^{ip_1\cdot X_1}e^{i p_2\cdot X_2}\rangle \langle p_1\cdot\psi_1\psi_1^\mu p_2\cdot\bar{\psi}_2\bar{\psi}_2^\sigma\rangle\,,
\end{equation}
\begin{equation}
\mathfrak{C}_4^{(4)}:=-g_c^2f_{\mu\nu}^1f_{\rho\sigma}^2\langle \bar{\partial} X_1^\nu\bar{\partial} X_2^\sigma e^{ip_1\cdot X_1}e^{i p_2\cdot X_2}\rangle \langle p_1\cdot\psi_1\psi_1^\mu p_2\cdot\psi_2\psi_2^\rho\rangle\,,
\end{equation}
we find
\begin{equation}\label{eqn:four ferm c2 0}
\mathfrak{C}_{4,e}^{(2)}=2g_c^2(p_1\cdot p_2)^2 \langle \partial X_1 X_2\rangle \langle X_1\bar{\partial} X_2\rangle \langle \bar{\psi}_1\psi_2\rangle^2e^{-p_1\cdot p_2\langle X_1X_2\rangle}\,,
\end{equation}
\begin{equation}\label{eqn:four ferm c3 0}
\mathfrak{C}_{4,e}^{(3)}=2g_c^2(p_1\cdot p_2)^2 \langle \bar{\partial} X_1 X_2\rangle \langle X_1\partial X_2\rangle \langle \psi_1\bar{\psi}_2\rangle^2e^{-p_1\cdot p_2\langle X_1X_2\rangle}\,,
\end{equation}
\begin{equation}\label{eqn:four ferm c4 0}
\mathfrak{C}_{4,e}^{(2)}=2g_c^2(p_1\cdot p_2)^2 \langle \bar{\partial} X_1 X_2\rangle \langle X_1\bar{\partial} X_2\rangle \langle \psi_1\psi_2\rangle^2e^{-p_1\cdot p_2\langle X_1X_2\rangle}\,.
\end{equation}

Because, as we stressed before, although the equations \eqref{eqn:four ferm c1 0}-\eqref{eqn:four ferm c4 0} are naively order of $\mathcal{O}(\delta^2),$ integral over the vertex moduli around the poles can bring down the factor of $1/\delta.$ We shall now carefully study this vertex collision.

Let us start with \eqref{eqn:four ferm c1 0}. Let us recall the identities
\begin{equation}
\langle \bar{\psi}_1\bar{\psi}_2\rangle^2=\biggl(-G_F(I_\sigma(z_1),I_\sigma(z_2);\tau,s)\biggr)^2\,,
\end{equation}
and
\begin{equation}
G_F(\bar{z},0;\tau,s)^2=\left(\frac{\vartheta_{\alpha,\beta}(\bar{z}|\tau)\vartheta_{1,1}'(0|\tau)}{\vartheta_{\alpha,\beta}(0|\tau)\vartheta_{1,1}(\bar{z}|\tau)}\right)^2=\frac{\vartheta_{\alpha,\beta}''(0|\tau)}{\vartheta_{\alpha,\beta}(0|\tau)}-\partial_{\bar{z}}^2 \log \vartheta_{1,1}(\bar{z}|\tau)\,.
\end{equation}
Because we are working with the even spin structure, we need to sum over the even spin structure. As the term
\begin{equation}
\partial_{\bar{z}}^2\log\vartheta_{1,1}(\bar{z}|\tau)
\end{equation}
does not depend on the spin structure, upon the spin sum this term will drop out. So, effectively we can treat the fermion correlator as
\begin{equation}
\langle \bar{\psi}_1\bar{\psi}_2\rangle^2=\frac{\vartheta_{\alpha,\beta}''(0|\tau)}{\vartheta_{\alpha,\beta}(0|\tau)}\,,
\end{equation}
which does not depend on the vertex moduli. Now, let us evaluate the bosonic correlators. We have
\begin{equation}
\langle \partial X_1(z_1) X_2(z_2)\rangle= -\frac{\alpha'}{2} \left(\frac{\vartheta_{1,1}'(z_1-z_2|\tau)}{\vartheta_{1,1}(z_1-z_2|\tau)}+\frac{\vartheta_{1,1}'(z_1-I_\sigma(z_2)|\tau)}{\vartheta_{1,1}(z_1-I_\sigma(z_2)|\tau)}\right)+\dots\,,
\end{equation}
\begin{equation}
\langle  X_1(z_1) \partial X_2(z_2)\rangle= -\frac{\alpha'}{2} \left(\frac{\vartheta_{1,1}'(z_2-z_1|\tau)}{\vartheta_{1,1}(z_2-z_1|\tau)}+\frac{\vartheta_{1,1}'(z_2-I_\sigma(z_1)|\tau)}{\vartheta_{1,1}(z_2-I_\sigma(z_1)|\tau)}\right)+\dots\,,
\end{equation}
where $\dots$ denotes non-singular terms. As a result, collecting only the relevant terms, we find
\begin{align}\label{eqn: four ferm c1 3}
\mathfrak{C}_{4,e}^{(1)}=&g_c^2 \frac{\delta^2}{2}\frac{\vartheta_{\alpha,\beta}''(0|\tau)}{\vartheta_{\alpha,\beta}(0|\tau)}\left(\frac{\vartheta_{1,1}'(z_1-z_2|\tau)}{\vartheta_{1,1}(z_1-z_2|\tau)}+\frac{\vartheta_{1,1}'(z_1-I_\sigma(z_2)|\tau)}{\vartheta_{1,1}(z_1-I_\sigma(z_2)|\tau)}\right)\nonumber\\
&\times\left(\frac{\vartheta_{1,1}'(z_2-z_1|\tau)}{\vartheta_{1,1}(z_2-z_1|\tau)}+\frac{\vartheta_{1,1}'(z_2-I_\sigma(z_1)|\tau)}{\vartheta_{1,1}(z_2-I_\sigma(z_1)|\tau)}\right)e^{-p_1\cdot p_2 (G_B(z_1,z_2)+G_B(z_1,I_\sigma(z_2)))}\,.
\end{align}

Before we evaluate \eqref{eqn: four ferm c1 3}, let us first take a moment to study what types poles we need to look for. To bring down the factor of $1/\delta,$ we shall find the following form of the pole
\begin{equation}
\delta |\delta z|^{\delta-2}\,,
\end{equation}
for small $\delta z.$ The reason is rather simple. Because only the local region around $\delta z=0$ is important, we can focus on integral over $\delta z$ around a small disk of radius $\epsilon$ such that $|\delta z|\leq \epsilon.$ We shall first let $\delta$ to be small but positive. Within the small disk, we can parametrize
\begin{equation}
\delta z= re^{i\theta}\,,
\end{equation}
so that the volume integral within the small disk is written as
\begin{equation}
\delta \int d^2\delta z |\delta z|^{\delta-2}= 2\delta \int_0^{2\pi}\int_0^{\epsilon} r^{\delta-1}dr d\theta=4\pi \epsilon^{\delta}\,.
\end{equation}
Because the exponential function can be analytically continues, we can now safely take $\delta\rightarrow 0$ limit to produce $1/\delta$ enhancement. 

In \eqref{eqn: four ferm c1 3}, the only relevant pole is therefore $z_1=I_\sigma (z_2).$ The pole from $z_1=z_2$ will yield zero contribution. We shall parametrize the vertex moduli using
\begin{equation}
\delta z:= z_1-I_\sigma(z_2)\,,
\end{equation}
and $z_2.$ The singular term around $\delta z=0,$ is given as
\begin{align}
\mathfrak{C}_{4,e}^{(1)} =&g_c^2\frac{\delta^2}{2}\frac{\vartheta_{\alpha,\beta}''(0|\tau)}{\vartheta_{\alpha,\beta}(0|\tau)}\frac{\vartheta_{1,1}'(z_1-I_\sigma(z_2)|\tau)}{\vartheta_{1,1}(z_1-I_\sigma(z_2)|\tau)}\frac{\vartheta_{1,1}'(z_2-I_\sigma(z_1)|\tau)}{\vartheta_{1,1}(z_2-I_\sigma(z_1)|\tau)}e^{-p_1\cdot p_2 G_B(z_1,I_\sigma(z_2))}+\dots\,,\\
=&g_c^2 \frac{\delta^2}{2} \frac{\vartheta_{\alpha,\beta}''(0|\tau)}{\vartheta_{\alpha,\beta}(0|\tau)} |\delta z|^{\delta -2}+\dots\,,
\end{align}
where we used
\begin{equation}
\frac{\vartheta_{1,1}'(z_1-I_\sigma(z_2)|\tau)}{\vartheta_{1,1}(z_1-I_\sigma(z_2)|\tau)}=\frac{1}{\delta z}+\mathcal{O}(\delta z)\,,
\end{equation}
and
\begin{equation}
\frac{\vartheta_{1,1}'(z_2-I_\sigma(z_1)|\tau)}{\vartheta_{1,1}(z_2-I_\sigma(z_1)|\tau)}=\frac{1}{\overline{\delta z}}+\mathcal{O}(\overline{\delta z})\,.
\end{equation}
Note that to arrive at the above mentioned equations we used
\begin{equation}
I_\sigma(z_1)=I_\sigma(I_\sigma(z_2)+\delta z)=z_2-\overline{\delta z}+\delta_{\sigma\mathcal{K}}\tau_\mathcal{K}\,.
\end{equation}
As a result, we find that upon the integral over $\delta z,$ $\mathfrak{C}_{4,e}^{(1)}$ produces the following term of order $\mathcal{O}(\delta)$ in the small $\delta$ limit 
\begin{equation}
\int_{|\delta z|\leq \epsilon} d^2\delta z\mathfrak{C}_{4,e}^{(1)}=2\pi\delta g_c^2 \frac{\vartheta_{\alpha,\beta}''(0|\tau)}{\vartheta_{\alpha,\beta}(0|\tau)}\,.
\end{equation}
Note that the above term is absent in the torus amplitude, as the bosonic correlator simply does not contain a term that has a pole at $z_1=I_\sigma (z_2).$ 

We shall now study \eqref{eqn:four ferm c2 0}. Let us recall the identity
\begin{equation}
\langle \bar{\psi}_1 \psi_2\rangle^2=\biggl(iG_F(I_\sigma(z_1),z_2;\tau,s)\biggr)^2\,.
\end{equation}
Note that this fermion correlator vanishes for the torus amplitude. Same as before, we can treat the fermion correlator as
\begin{equation}
\langle \bar{\psi}_1 \psi_2\rangle^2=-\frac{\vartheta_{\alpha,\beta}''(0|\tau)}{\vartheta_{\alpha,\beta}(0|\tau)}\,,
\end{equation}
because the $\partial^2\log \vartheta_{1,1}(z|\tau)$ term does not depend on the spin structure. Note the additional negative sign due to the $i$ factor in the fermionic correlator.  Now, let us evaluate the bosonic correlators. We have
\begin{equation}
\langle \partial X_1(z_1) X_2(z_2)\rangle= -\frac{\alpha'}{2} \left(\frac{\vartheta_{1,1}'(z_1-z_2|\tau)}{\vartheta_{1,1}(z_1-z_2|\tau)}+\frac{\vartheta_{1,1}'(z_1-I_\sigma(z_2)|\tau)}{\vartheta_{1,1}(z_1-I_\sigma(z_2)|\tau)}\right)+\dots\,,
\end{equation}
\begin{equation}
\langle  X_1(z_1) \bar{\partial} X_2(z_2)\rangle= -\frac{\alpha'}{2} \left(\frac{\vartheta_{1,1}'(\bar{z}_2-\bar{z}_1|\tau)}{\vartheta_{1,1}(\bar{z}_2-\bar{z}_1|\tau)}+\frac{\vartheta_{1,1}'(\bar{z}_2-I_\sigma(\bar{z}_1)|\tau)}{\vartheta_{1,1}(\bar{z}_2-I_\sigma(\bar{z}_1)|\tau)}\right)+\dots\,,
\end{equation}
where $\dots$ denotes non-singular terms. Collecting all those correlators, we find
\begin{align}\label{eqn: four ferm c2 1}
\mathfrak{C}_{4,e}^{(2)}=&-g_c^2 \frac{\delta^2}{2}\frac{\vartheta_{\alpha,\beta}''(0|\tau)}{\vartheta_{\alpha,\beta}(0|\tau)}\left(\frac{\vartheta_{1,1}'(z_1-z_2|\tau)}{\vartheta_{1,1}(z_1-z_2|\tau)}+\frac{\vartheta_{1,1}'(z_1-I_\sigma(z_2)|\tau)}{\vartheta_{1,1}(z_1-I_\sigma(z_2)|\tau)}\right)\nonumber\\
&\times\left(\frac{\vartheta_{1,1}'(\bar{z}_2-\bar{z}_1|\tau)}{\vartheta_{1,1}(\bar{z}_2-\bar{z}_1|\tau)}+\frac{\vartheta_{1,1}'(\bar{z}_2-I_\sigma(\bar{z}_1)|\tau)}{\vartheta_{1,1}(\bar{z}_2-I_\sigma(\bar{z}_1)|\tau)}\right)e^{-p_1\cdot p_2 (G_B(z_1,z_2)+G_B(z_1,I_\sigma(z_2)))}\,.
\end{align}

In \eqref{eqn: four ferm c2 1}, the only relevant pole is from
\begin{equation}
z_1=z_2\,.
\end{equation}
Note that the pole from 
\begin{equation}
z_1=I_\sigma (z_2)\,,
\end{equation}
does not generate a non-vanishing contribution. Let us therefore study the pole from $z_1=z_2.$ We shall define $\delta z$ as
\begin{equation}
\delta z:=z_1-z_2\,.
\end{equation}
We collect the most singular terms in $\delta z$ in \eqref{eqn: four ferm c2 1}
\begin{equation}
\mathfrak{C}_{4,e}^{(2)}= g_c^2 \frac{\delta^2}{2} \frac{\vartheta_{\alpha,\beta}''(0|\tau)}{\vartheta_{\alpha,\beta}(0|\tau)}|\delta z|^{\delta-2}+\dots\,.
\end{equation}
Therefore, we find
\begin{equation}
\int_{|\delta z|\leq \epsilon} \mathfrak{C}_{4,e}^{(2)}=2\pi \delta g_c^2\frac{\vartheta_{\alpha,\beta}''(0|\tau)}{\vartheta_{\alpha,\beta}(0|\tau)}\,.
\end{equation}

To show that the pole from $z_1=I_\sigma(z_2)$ does not yield a non-trivial result, let us look at the pole from $z_1=I_\sigma(z_2)$ more closely. Let $\delta z'$ to be defined as
\begin{equation}
\delta z':=z_1-I_\sigma(z_2)\,.
\end{equation}
Then, we have
\begin{equation}
\frac{\vartheta_{1,1}'(z_1-I_\sigma(z_2)|\tau)}{\vartheta_{1,1}(z_1-I_\sigma(z_2)|\tau)}=\frac{1}{\delta z'}+\dots\,,
\end{equation}
and
\begin{equation}
\frac{\vartheta_{1,1}'(\bar{z}_2-I_\sigma(\bar{z}_1)|\tau)}{\vartheta_{1,1}(\bar{z}_2-I_\sigma(\bar{z}_1)|\tau)}=\frac{1}{\delta z'}+\dots\,.
\end{equation}
Note that we used
\begin{equation}
I_\sigma(\bar{z}_1)=I_\sigma(\overline{I_\sigma(z_2)}+\overline{\delta z}')=I_\sigma\left(1-z_2+\overline{\delta z}'-\frac{\delta_{\sigma\mathcal{K}}}{2}\tau_\mathcal{K}\right)=\bar{z}_2-\delta z'\,.
\end{equation}
Collecting the most singular terms, we find
\begin{equation}
\mathfrak{C}_{4,e}^{(2)}= g_c^2 \frac{\delta^2}{2} \frac{\vartheta_{\alpha,\beta}''(0|\tau)}{\vartheta_{\alpha,\beta}(0|\tau)}\frac{|\delta z'|^{\delta}}{\delta z'^2}+\dots\,.
\end{equation}
Therefore, we find
\begin{equation}
\int_{|\delta z'|\leq \epsilon} \mathfrak{C}_{4,e}^{(2)}=0\,.
\end{equation}
Note that we used
\begin{equation}
\int_{|\delta z'|\leq\epsilon}d^2( \delta z')\left[ \delta z'^{\delta/2-2}\overline{\delta z}'^{\delta/2}\right]=0\,.
\end{equation}
As a result, we find that from \eqref{eqn:four ferm c2 0} we obtain
\begin{equation}
2\pi\delta g_c^2\frac{\vartheta_{\alpha,\beta}''(0|\tau)}{\vartheta_{\alpha,\beta}(0|\tau)}\,.
\end{equation}

The contribution from \eqref{eqn:four ferm c3 0} can be read off from by replacing $z_1$ with $z_2$ in \eqref{eqn:four ferm c2 0}. Similarly, the contribution from \eqref{eqn:four ferm c4 0} can be read off by replaing $z_i$ with $\bar{z}_i$ in \eqref{eqn:four ferm c1 0}. Therefore, collecting all terms, after integrating over the pole region in the vertex moduli space, we find
\begin{equation}\label{eqn:four ferm res}
\int_{|\delta z|\leq \epsilon} \mathfrak{C}_{4,e}=8\pi\delta g_c^2\frac{\vartheta_{\alpha,\beta}''(0|\tau)}{\vartheta_{\alpha,\beta}(0|\tau)}\,.
\end{equation}
It is important to note that for the torus amplitude there is no non-trivial contribution as we stressed before in this section.


\subsubsection{Eight fermion terms: genuine contractions}
We shall now study the eight fermion contributions. We write
\begin{equation}
\mathfrak{C}_8=\frac{\alpha'^2}{4} f^1_{\mu\nu}f^2_{\rho\sigma}\left\langle p_1\cdot\psi_1\psi_1^\mu p_1\cdot \bar{\psi}_1\bar{\psi}_1^\nu p_2\cdot \psi_2 \psi_2^\rho p_2\cdot\bar{\psi}_2\bar{\psi}_2^\sigma\right\rangle\,.
\end{equation}
There are in total 24 different genuine contractions one can find. We will start by studying two different contractions, and we will permute the contractions to obtain the full answer. 

We first consider
\begin{align}
\mathfrak{C}_8^{(1)}:=&\frac{\alpha'^2}{4}f^1_{\mu\nu}f^2_{\rho\sigma}\langle p_1\cdot \psi_1  p_2\cdot \psi_2\rangle\langle \psi_1^\mu \psi_2^\rho\rangle\langle p_1\cdot\bar{\psi}_1 p_2\cdot \bar{\psi}_2 \rangle\langle \bar{\psi}_1^\nu\bar{\psi}_2^\sigma\rangle\,,\\
=&\frac{\alpha'^2}{4} f_{\mu\nu}^1f_{\rho\sigma}^2 (p_1\cdot p_2)^2 \eta^{\mu\rho}\eta^{\nu\sigma} |(\langle \psi_1\psi_2\rangle)|^4\,.
\end{align}
Next, we then consider
\begin{align}
\mathfrak{C}_8^{(2)}:=&-\frac{\alpha'^2}{4}f_{\mu\nu}^1f_{\rho\sigma}^2 \langle p_1\cdot \psi_1 \psi_2^\rho\rangle \langle \psi_1^\mu p_2\cdot \psi_2\rangle \langle p_1\cdot \bar{\psi}_1 p_2\cdot \bar{\psi}_2\rangle \langle \bar{\psi}_1^\nu \bar{\psi}_2^\sigma\rangle\,,\\
=& -\frac{\alpha'^2}{4} f^1_{\mu\nu}f^2_{\rho\sigma} p_1^\rho p_2^\mu p_1\cdot p_2 \eta^{\rho\sigma} |\langle \psi_1\psi_2\rangle|^4\,.
\end{align}
And, same as in the four fermion contractions, we find 
\begin{equation}
    \mathfrak{C}_8^{(1)}+\mathfrak{C}_8^{(2)}=0\,.
\end{equation}
By permuting the contractions, one can find that 8 terms out of 24 contractions cancel out due to the relative signs, and 8 terms cancel due to the kinematic relation as above
\begin{equation}
f_{\mu\nu}^1f_{\rho\sigma}^2 (p_1\cdot p_2)\eta^{\nu\sigma}(\eta^{\mu\rho}-p_1^\rho p_2^\mu)=0\,,
\end{equation}
and the rest of the terms survive. We simplify the rest of the terms using the kinematic relation
\begin{equation}
f_{\mu\nu}^1f_{\rho\sigma}^2 p_1^\rho p_2^\mu (\eta^{\nu\sigma}-p_1^\sigma p_2^\nu)=-2(p_1\cdot p_2)^2\,,
\end{equation}
into
\begin{equation}\label{eqn:G intermediate}
\mathcal{G}=\frac{1}{2}(\alpha'p_1\cdot p_2)^2 (G_F(z_1,I_\sigma(z_2))G_F(I_\sigma (z_1),z_2)- G_F(z_1,z_2)G_F(I_\sigma (z_1),I_\sigma (z_2)))^2\,.
\end{equation}
We will show the detailed demonstration of the computation of \eqref{eqn:G intermediate} in \S\ref{app:cancellation}.

We divide $\mathcal{G}$ into three different terms $\mathcal{G}=\mathcal{G}_1+\mathcal{G}_2+\mathcal{G}_3$
\begin{align}
\mathcal{G}_1=&\frac{\delta^2}{2} \biggl(G_F(z_1,I_\sigma(z_2)) G_F(I_\sigma(z_1),z_2) \biggr)^2\,,\\
\mathcal{G}_2=&-\delta^2 G_F(z_1,I_\sigma(z_2))G_F(I_\sigma (z_1),z_2) G_F(z_1,z_2)G_F(I_\sigma (z_1),I_\sigma (z_2))\,,\\
\mathcal{G}_3=&\frac{\delta^2}{2} \biggl( G_F(z_1,z_2)G_F(I_\sigma (z_1),I_\sigma (z_2))\biggr)^2\,.
\end{align}
We will study how vertex collisions can bring down a factor of $1/\delta$ for each case one by one.

Let us start with $\mathcal{G}_1.$ We shall use the identities
\begin{equation}
G_F(z_1,I_\sigma(z_2);\tau,s)^2= \frac{\vartheta_{\alpha,\beta}''(0|\tau)}{\vartheta_{\alpha,\beta}(0|\tau)}-\partial_z^2 \log \vartheta_{1,1}(z|\tau)|_{z=z_1-I_\sigma(z_2)}\,,
\end{equation}
and
\begin{equation}
G_F(I_\sigma(z_1),z_2;\tau,s)^2= \frac{\vartheta_{\alpha,\beta}''(0|\tau)}{\vartheta_{\alpha,\beta}(0|\tau)}-\partial_z^2 \log \vartheta_{1,1}(z|\tau)|_{z=I_\sigma(z_1)-z_2}\,.
\end{equation}
Because the vertex position dependence only comes from $\partial_z^2 \log\vartheta_{1,1}(z|\tau)$ term, the following term
\begin{equation}
\mathcal{G}_1\supset\frac{\delta^2}{2}\left(\frac{\vartheta_{\alpha,\beta}''(0|\tau)}{\vartheta_{\alpha,\beta}(0|\tau)}\right)^2
\end{equation}
does not have a pole, and therefore it cannot produce $\mathcal{O}(\delta)$ term upon the integral over the vertex operator position. On the other hand, the following term
\begin{equation}\label{eqn:G1 2}
\mathcal{G}_1\supset \frac{\delta^2}{2}\partial_z^2 \log \vartheta_{1,1}(z|\tau)|_{z=z_1-I_\sigma(z_2)}\partial_z^2 \log \vartheta_{1,1}(z|\tau)|_{z=I_\sigma(z_1)-z_2}
\end{equation}
does not depend on the spin structure. Therefore, the contribution from \eqref{eqn:G1 2} vanishes upon the sum over the spin structures. Therefore, the only non-trivial contribution at order $\mathcal{O}(\delta)$ can possibly come from
\begin{equation}\label{eqn:G1 3}
\mathcal{G}_1 \supset -\frac{\delta^2}{2} \frac{\vartheta_{\alpha,\beta}''(0|\tau)}{\vartheta_{\alpha,\beta}(0|\tau)}\biggl[\partial_z^2 \log \vartheta_{1,1}(z|\tau)|_{z=z_1-I_\sigma(z_2)}+\partial_z^2 \log \vartheta_{1,1}(z|\tau)|_{z=I_\sigma(z_1)-z_2}\biggr]\,.
\end{equation}

Now, we shall show that in fact even \eqref{eqn:G1 3} does not produce a non-trivial result at order $\mathcal{O}(\delta)$ under the integral over the vertex position. It is important to note that \eqref{eqn:G1 3} has a pole when $z_1\rightarrow I_\sigma(z_2).$ To understand when such a vertex collision can produce $\mathcal{O}(\delta)$ contribution, we shall also at the same time consider the bosonic correlator
\begin{equation}\label{eqn:bosonic corr sec4}
\mathcal{X}:=\left\langle e^{ip_1\cdot X_1}e^{ip_2\cdot X_2}\right\rangle=\exp\left[-p_1\cdot p_2 \left(G_B(z_1,z_2;\tau)+G_B(z_1,I_\sigma (z_2);\tau)\right)\right]\,.
\end{equation}

Let define $\delta z:= z_1-I_\sigma (z_2).$ Because the factor $1/\delta$ can arise under integrating over $\delta z$ for small $\delta z,$ it suffices to consider a small disk of radius $\epsilon,$ $D_\epsilon,$ i.e. $|\delta z|\leq \epsilon.$  When $\delta z$ is small, the bosonic correlator can be well approximated as
\begin{equation}
\left\langle e^{ip_1\cdot X_1}e^{ip_2\cdot X_2}\right\rangle=|\delta z|^{\delta}\exp\left[-p_1\cdot p_2 G_B(z_2,I_\sigma(z_2);\tau)\right]+\dots\,,
\end{equation}
For small $\delta z,$ we have
\begin{equation}
\partial_z^2\log\vartheta_{1,1}(z|\tau)|_{z=z_1-I_\sigma (z_2)}=-\frac{1}{\delta z^2}+\dots\,.
\end{equation}
So, effectively, to understand the contribution from the first term in \eqref{eqn:G1 3}, we need to compute the following integral 
\begin{equation}\label{eqn:G1 4}
\int_{|\delta z|\leq \epsilon} d^2\delta z \biggl(\delta z^{\delta/2-2}\overline{\delta z}^{\delta/2} \biggr)\,.
\end{equation}
We rewrite the variable $\delta z$ as
\begin{equation}
\delta z= r e^{i\theta}\,,
\end{equation}
where $0\leq r\leq \epsilon,$ and $0\leq\theta\leq2\pi,$ to convert the integral \eqref{eqn:G1 4} into
\begin{equation}
2\int_0^{2\pi}\int_0^\epsilon r^{\delta-1} e^{-2i\theta} dr d\theta\,,
\end{equation}
which vanishes due to 
\begin{equation}
\int_0^{2\pi} e^{-2i\theta}d\theta=0\,.
\end{equation}
As a result, we conclude that the first term in \eqref{eqn:G1 3} does not yield any contribution to order $\mathcal{O}(\delta).$ Similarly, we conclude that the second term in \eqref{eqn:G1 3} and $\mathcal{G}_3$ also do not yield any contribution at order $\mathcal{O}(\delta).$ 

We shall now study $\mathcal{G}_2.$ There are two different poles: one from $z_1=I_\sigma(z_2),$ and the other from $z_1=z_2.$ 

Let us first study the pole from $z_1\rightarrow I_\sigma(z_2).$ We define $\delta z:= z_1-I_\sigma(z_2)$ as before. Then we have
\begin{equation}
I_\sigma(z_1)=I_\sigma(\delta z+I_\sigma (z_2))=z_2-\overline{\delta z}+\delta_{\sigma \mathcal{K}}\tau\,.
\end{equation}
Note that if the worldsheet is Klein bottle, we have an additional term $\tau.$ Same as before, we will zoom in to the small disk of radius of $\epsilon$ around $\delta z=0,$ $D_\epsilon.$ When $\delta z$ is small, we can approximate
\begin{equation}
G_F(z_1,I_\sigma(z_2);\tau,s)= \frac{1}{\delta z}+\mathcal{O}(\delta z)\,,
\end{equation}
and
\begin{equation}
G_F(I_\sigma(z_1),z_2;\tau,s)=-\frac{e^{-i\pi(\beta-1)\delta_{\sigma\mathcal{K}}}}{\overline{\delta z}}+\mathcal{O}(\overline{\delta z})\,.
\end{equation}
Then, the leading term in $\delta z$ is given as
\begin{equation}\label{eqn:G2 vert1 1}
\mathcal{G}_2= -\delta^2 \frac{1}{|\delta z|^2} \biggl(G_F(z_2,I_\sigma(z_2))\biggr)^2+\dots\,,
\end{equation}
where $\dots$ denotes the higher order terms in $\delta z.$ Note that we used the following identity
\begin{align}
G_F(I_\sigma(z_1),I_\sigma(z_2))=&G_F(z_2+\delta_{\sigma\mathcal{K}}\tau,I_\sigma(z_2))+\dots\,\\
=&e^{-i\pi (\beta-1)\delta_{\sigma\mathcal{K}}}G_F(z_2,I_\sigma(z_2))+\dots\,.
\end{align}
Now, combining \eqref{eqn:G2 vert1 1} with the bosonic correlator \eqref{eqn:bosonic corr sec4} when $\delta z$ is small, we have
\begin{equation}
\mathcal{G}_2 \mathcal{X}=-\delta^2 |\delta z|^{\delta-2}\biggl(G_F(z_2,I_\sigma(z_2))\biggr)^2\exp\left[-p_1\cdot p_2 G_B(z_2,I_\sigma(z_2);\tau)\right]+\dots\,.
\end{equation}

We shall now perform the integral over $\delta z.$ To make the integral well behaving, we will first let $\delta=\alpha' p_1\cdot p_2$ to be positive, and we will then take the small $\delta$ limit after performing an analytic continuation.  We first note that the integral
\begin{equation}
\frac{1}{\delta}\int_{|\delta z|\leq\epsilon} d^2\delta z |\delta z|^{\delta-2}
\end{equation}
over the small disk $D_\epsilon$ can be rewritten as
\begin{equation}\label{eqn:pole integral}
\frac{4\pi}{\delta}\int_0^\epsilon dr r^{\delta-1}=4\pi \epsilon^\delta\,.  
\end{equation}
Because any exponential function admits a series expansion with infinite radius of convergence, we can now perform an analytic continuation of \eqref{eqn:pole integral}, and take the small $\delta$ limit. As a result, we have
\begin{equation}\label{eqn:pole integral2}
\lim_{\delta\rightarrow 0}\frac{1}{\delta}\int_{|\delta z|\leq\epsilon}d^2\delta z |\delta z|^{\delta-2}=4\pi\,.
\end{equation}

By using \eqref{eqn:pole integral2}, in the small $\delta$ limit, we compute
\begin{equation}\label{eqn:res G2 1}
\int_{|\delta z|\leq\epsilon} d^2\delta z \mathcal{G}_2 \mathcal{X}=-2\pi \delta \biggl(G_F(z_2,I_\sigma(z_2))\biggr)^2\exp\left[-p_1\cdot p_2 G_B(z_2,I_\sigma(z_2);\tau)\right]\,.
\end{equation}

Now, let us study the contribution from thepole from $z_1\rightarrow z_2$ in $\mathcal{G}_2.$ This time, we shall define $\delta z':= z_1-z_2.$ We will, again, restrict to small disk of radius $\epsilon,$ $D_\epsilon'.$ When $\delta z'$ is small, we can approximate
\begin{equation}
G_F(z_1,z_2)=\frac{1}{\delta z'}+\mathcal{O}(\delta z)\,,
\end{equation}
and
\begin{equation}
G_F(I_\sigma(z_1),I_\sigma(z_2))=-\frac{1}{\overline{\delta z'}}+\mathcal{O}(\overline{\delta z})\,.
\end{equation}
As a result, the leading term in $\delta z'$ is given as
\begin{equation}\label{eqn:G2 vert1 2}
\mathcal{G}_2=-\delta^2 \frac{1}{|\delta z'|^2} \biggl(G_F(z_2,I_\sigma(z_2);\tau,s)\biggr)^2+\dots\,,
\end{equation}
where $\dots$ represents higher order terms in $\delta z'.$ By combining \eqref{eqn:G2 vert1 2} with \eqref{eqn:bosonic corr sec4}, in the small $\delta z$ limit, we find
\begin{equation}
\mathcal{G}_2\mathcal{X}=-\delta^2 |\delta z'|^{\delta-2}\left[-p_1\cdot p_2 G_B(z_2,I_\sigma(z_2);\tau)\right]+\dots\,.
\end{equation}
By performing the integral over $\delta z',$ in the small $\delta$ limit, we again find
\begin{equation}\label{eqn:res G2 2}
\int_{|\delta z'|\leq\epsilon}d^2\delta z' \mathcal{G}_2 \mathcal{X}=-2\pi \delta \biggl(G_F(z_2,I_\sigma(z_2))\biggr)^2\exp\left[-p_1\cdot p_2 G_B(z_2,I_\sigma(z_2);\tau)\right]\,.
\end{equation}

Combining \eqref{eqn:res G2 1} and \eqref{eqn:res G2 2} we find that the genuine contractions, at order $\mathcal{O}(\delta),$ produce the contribution
\begin{equation}\label{eqn:genuine net}
-8\pi \delta \biggl(G_F(z_2,I_\sigma(z_2))\biggr)^2\exp\left[-p_1\cdot p_2 G_B(z_2,I_\sigma(z_2);\tau)\right]\,.
\end{equation}
\subsubsection{Eight fermion terms: self-contractions}
Now, we shall compute the self-contraction contributions. Because of the mass-shell conditions
\begin{equation}
p_1^2=p_2^2=0\,,
\end{equation}
and the transversality conditions for the polarizations
\begin{equation}
f_{\mu\nu}^1p^\mu_1=f_{\mu\nu}^2p_2^\mu=0\,,
\end{equation}
most of the self-contractions vanish. But, there are two self-contractions that yield non-trivial results
\begin{equation}
\mathcal{S}_1:=\frac{\alpha'^2}{4} f_{\mu\nu}^1f_{\rho\sigma}^2 \langle p_1 \cdot \bar{\psi}_1 p_2\cdot \psi_2\rangle \langle p_1\cdot \psi_1 p_2\cdot \bar{\psi}_2\rangle \langle \psi_1^\mu\bar{\psi}_1^\nu\rangle \langle \psi_2^\rho\bar{\psi}_2^\sigma\rangle\,,
\end{equation}
and
\begin{equation}
\mathcal{S}_2:=-\frac{\alpha'^2}{4} f_{\mu\nu}^1f_{\rho\sigma}^2\langle p_1\cdot\psi_1 p_2\cdot \psi_2 \rangle\langle p_1\cdot \bar{\psi}_1 p_2\cdot \bar{\psi}_2\rangle\langle \psi_1^\mu\bar{\psi}_1^\nu\rangle \langle \psi_2^\rho\bar{\psi}_2^\sigma\rangle\,.
\end{equation}
Above two contractions vanish for the torus diagram, but need not vanish for annuli, M\"{o}bius strip, and Klein bottle. Using the Green's functions summarized in \S\ref{sec:Green function}, we compute
\begin{equation}
\mathcal{S}_1=\delta^2 G_F(I_\sigma (z_1), z_2)G_F(z_1,I_\sigma (z_2) )G_F(z_1,I_\sigma (z_1))G_F(z_2,I_\sigma (z_2))\,,
\end{equation}
and
\begin{equation}
\mathcal{S}_2=-\delta^2G_F(z_1,z_2)G_F(I_\sigma (z_1),I_\sigma (z_2))G_F(z_1,I_\sigma (z_1))G_F(z_2,I_\sigma (z_2))\,.
\end{equation}
Note here that we supressed the spin index for the fermion correlators.

To understand when $S_1$ and $S_2$ contribute to the amplitude at order $\mathcal{O}(p_1\cdot p_2),$ we need to study the pole structure of the fermion correlators. Because the contractions between $e^{ip_1\cdot X_1}$ and $e^{ip_2\cdot X_2}$ generate the factor
\begin{equation}
\langle e^{ip_1\cdot X_1}e^{ip_2\cdot X_2}\rangle= \exp\left[-p_1\cdot p_2 \left(G_B(z_1,z_2;\tau)+G_B(z_1,I_\sigma (z_2);\tau)\right)\right]\,,
\end{equation}
if there are poles of the form $|z_1-z_2|^{-2}$ or $|z_1-I_\sigma(z_2)|^{-2},$ upon the vertex moduli integral, one can generate $-1/(p_1\cdot p_2).$ Therefore, we shall closely examine the pole structures and the vertex moduli integral around the poles. Note that the self-contractions for bosonic fields vanish because of the mass-shell conditions. 

Let us first study $\mathcal{S}_1.$ The relevant pole is when $z_1=I_\sigma (z_2).$ We will therefore parametrize the vertex moduli using $z_2$ and $\delta z,$ which we define as
\begin{equation}
\delta z:= z_1-I_\sigma (z_2)\,.
\end{equation}
Because only the pole region contributes to the integral to produce $\mathcal{O}(\delta)$ contribution, we will restrict the domain of integral to a small disk of radius $\epsilon,$ $D_\epsilon$,  i.e. $|\delta z|\leq \epsilon.$ Then, we can approximate
\begin{align}
G_F(I_\sigma(z_1),z_2;\tau,s)=& G_F(z_2-\overline{\delta z}+\delta_{\sigma\mathcal{K}}\tau,z_2;\tau,s)\,\\
=&-\frac{e^{-i\pi(\beta-1)\delta_{\sigma\mathcal{K}}}}{\overline{\delta z}}+\mathcal{O}(\overline{\delta z})\,,
\end{align}
and
\begin{equation}
G_F(z_1,I_\sigma (z_2);\tau,s)=\frac{1}{\delta z}+\mathcal{O}(\delta z)\,.
\end{equation}
Similarly, we have
\begin{align}
\langle e^{ip_1\cdot X_1}e^{ip_2\cdot X_2}\rangle=&\exp\left[-p_1\cdot p_2 (G_B(I_\sigma (z_2+\delta z),z_2;\tau)+G_B(I_\sigma (z_2)+\delta z,I_\sigma(z_2);\tau) \right]\,,\\
=& |\delta z|^{\delta}\exp\left[-p_1\cdot p_2 G_B(I_\sigma (z_2),z_2;\tau)\right]+\dots\,.
\end{align}
Therefore, collecting the leading terms in $\delta z,$ we have
\begin{equation}
\mathcal{S}_1\mathcal{X}=\delta^2 \biggl(G_F(z_2,I_\sigma(z_2);\tau,s)\biggr)^2  |\delta z|^{\delta-2}\exp\left[-p_1\cdot p_2 G_B(I_\sigma (z_2),z_2;\tau)\right]+\dots\,.
\end{equation}
We again used
\begin{equation}
G_F(z_1,I_\sigma(z_1))=-e^{-i\pi (\beta-1)\delta_{\sigma\mathcal{K}}}G_F(z_2,I_\sigma(z_2))+\dots\,.
\end{equation}
By performing the integral over $\int_{|\delta z|\leq \epsilon} g_{\delta z\overline{\delta z}} d^2\delta z,$ using \eqref{eqn:pole integral2} we obtain 
\begin{equation}\label{eqn:vertex collision 1}
\int_{|\delta z|\leq \epsilon}  d^2\delta z \mathcal{S}_1\mathcal{X}=4\pi\delta\biggl(G_F(z_2,I_\sigma(z_2);\tau,s)\biggr)^2e^{-p_1\cdot p_2 G_B(I_\sigma (z_2),z_2;\tau)}\,.
\end{equation}

Let us now study $\mathcal{S}_2.$ The relevant pole is $z_1=z_2.$ We will therefore parametrize the vertex moduli using $z_2$ and $\delta z',$ where $\delta z'$ is now defined as
\begin{equation}
\delta z' := z_1-z_2\,.
\end{equation}
We will again focus on a small disk of radius $\epsilon,$ i.e. $|\delta z;|\leq \epsilon.$ Then, we find
\begin{equation}
G_F(z_1,z_2;\tau,s)= \frac{1}{\delta z'}+\mathcal{O}(\delta z)\,,
\end{equation}
and
\begin{equation}
G_F(I_\sigma (z_1),I_\sigma(z_2);\tau,s)=-\frac{1}{\overline{\delta z}'}+\mathcal{O}(\overline{\delta z})\,.
\end{equation}
Similarly, we have
\begin{equation}
\langle e^{ip_1\cdot X_1}e^{ip_2\cdot X_2}\rangle = |\delta z'|^{\alpha'p_1\cdot p_2} \exp\left[ -p_1\cdot p_2 G_B(z_2,I_\sigma(z_2);\tau)\right]+\dots\,.
\end{equation}
Collecting the leading terms in $\delta z,$ we find
\begin{equation}
\mathcal{S}_2\mathcal{X}=\delta^2 \biggl(G_F(z_2,I_\sigma(z_2);\tau,s)\biggr)^2|\delta z'|^{\delta-2}\exp\left[-p_1\cdot p_2 G_B(z_2,I_\sigma (z_2);\tau)\right]+\dots\,.
\end{equation}
Upon the integral over $\delta z',$ in small $\delta$ limit, we again find
\begin{equation}\label{eqn:vertex collision 2}
\int_{|\delta z'|\leq \epsilon} d^2\delta z' \mathcal{S}_2\mathcal{X}=4\pi\delta\biggl(G_F(z_2,I_\sigma(z_2);\tau,s)\biggr)^2e^{-p_1\cdot p_2 G_B(z_2,I_\sigma(z_2);\tau)}\,.
\end{equation}
Note that \eqref{eqn:vertex collision 1} and \eqref{eqn:vertex collision 2} have the same sign. By combining \eqref{eqn:vertex collision 1} and \eqref{eqn:vertex collision 2}, we therefore obtain
\begin{equation}\label{eqn:self net}
8\pi\delta  \biggl(G_F(z_2,I_\sigma(z_2);\tau,s)\biggr)^2e^{-p_1\cdot p_2 G_B(z_2,I_\sigma(z_2);\tau)}\,.
\end{equation}

\subsubsection{Results from the even spin structures}
Let us recall \eqref{eqn:genuine net}
\begin{equation}
-8\pi \delta \biggl(G_F(z_2,I_\sigma(z_2))\biggr)^2e^{-p_1\cdot p_2 G_B(z_2,I_\sigma(z_2);\tau)}\,,
\end{equation}
and \eqref{eqn:self net}
\begin{equation}
8\pi\delta  \biggl(G_F(z_2,I_\sigma(z_2);\tau,s)\biggr)^2e^{-p_1\cdot p_2 G_B(z_2,I_\sigma(z_2);\tau)}\,.
\end{equation}
Surprisingly, \eqref{eqn:genuine net} and \eqref{eqn:self net} completely cancel each other. The remaining term is given by \eqref{eqn:four ferm res} 
\begin{equation}
\int_{|\delta z|\leq \epsilon} \mathfrak{C}_{4,e}=8\pi\delta g_c^2\frac{\vartheta_{\alpha,\beta}''(0|\tau)}{\vartheta_{\alpha,\beta}(0|\tau)}\,.
\end{equation}

%

As a result, we can rewrite $Z_\sigma$ as 
\begin{equation}
Z_\sigma = \frac{c_\sigma V_4 (p_1\cdot p_2) g_c^2}{2^3\pi^3\alpha'}\sum_{\alpha,\beta \text{ even}}\int_0^\infty\frac{dt}{t^3}\int_\sigma d^2z_2 (-1)^{\alpha+\beta} \frac{\vartheta_{\alpha,\beta}''(\tau)}{\eta(\tau)^3}\text{Tr}_\alpha \left((-1)^{\beta F}q^{L_0-\frac{3}{8}} \right)^{\sigma}_{int}\,.
\end{equation}
After the integral over $d^2 z_2,$ which produces $\tau_{2\sigma},$ we therefore obtain
\begin{equation}
Z_\sigma = \frac{c_\sigma V_4 (p_1\cdot p_2) g_c^2}{2^3\pi^3\alpha'}\sum_{\alpha,\beta \text{ even}}\int_0^\infty\frac{dt}{t^3}\tau_{2\sigma} (-1)^{\alpha+\beta} \frac{\vartheta_{\alpha,\beta}''(\tau)}{\eta(\tau)^3}\text{Tr}_\alpha \left((-1)^{\beta F}q^{L_0-\frac{3}{8}} \right)^{\sigma}_{int}\,.
\end{equation}
Note that as is summarized in \S\ref{sec:Green function}, we have
\begin{equation}
\tau_A=it\,,\qquad \tau_{\mathcal{M}}=\frac{1}{2}+it\,,\qquad \tau_\mathcal{K}=2i t\,.
\end{equation}

As was studied in \cite{Kim:2023sfs}, the following identity holds
\begin{equation}
\sum_{\alpha,\beta\text{ even}} (-1)^{\alpha+\beta}\frac{\vartheta_{\alpha,\beta}''(\tau)}{4\pi^2\eta(\tau)^3}\text{Tr}_\alpha \left((-1)^{\beta F}q^{L_0-\frac{3}{8}} \right)^{\sigma}_{int}=-\text{Tr}_R\left((-1)^{F-\frac{3}{2}}Fq^{L_0-\frac{3}{8}}\right)^{\sigma}_{int}-\frac{3}{2}(n_\sigma^+-n_\sigma^-)\,,
\end{equation}
where $n^\pm_\sigma$ is a number of (anti)-BPS states in the partition function on $\sigma.$ We therefore arrive at
\begin{equation}
Z_\sigma=- \frac{c_\sigma V_4 (p_1\cdot p_2)g_c^2}{2 \pi\alpha'} \int_0^\infty\frac{dt}{t^3}\tau_{2\sigma} \left[\text{Tr}_R\left((-1)^{F-\frac{3}{2}}Fq^{L_0-\frac{3}{8}}\right)^{\sigma}_{int}+\frac{3}{2}(n_\sigma^+-n_\sigma^-)\right]\,.
\end{equation}
Finally, using the identification \eqref{eqn:identification}, we arrive at one of our main results
\begin{equation}
\boxed{\mathfrak{G}_{\sigma,\phi}^{(1)}=-\frac{c_\sigma}{2^3\pi^2} \int_0^\infty \frac{dt}{2t^3}\tau_{2\sigma}\left[\text{Tr}_R\left((-1)^{F-\frac{3}{2}}Fq^{L_0-\frac{3}{8}}\right)^{\sigma}_{int}+\frac{3}{2}(n_\sigma^+-n_\sigma^-)\right]\,.}
\end{equation}

It is interesting to note that $\mathfrak{G}_{\sigma,\phi\phi}^{(1)}$ does not suffer from the IR-divergence, as $\int \frac{dt}{t^2}$ converges as $t\rightarrow \infty.$ The naive UV divergence is cancelled by the tadpole cancellation.

\subsection{(odd,odd) spin structure}
Let us now consider the (odd,odd) spin structure. Contributions from such spin structures can only arise from closed string sectors: torus, Klein bottles. As the computation is mostly analogous, we shall focus on the torus contribution. In the (odd,odd) spin structure, due to the presence of the non-trivial conformal Killing vector we must fix $\theta$ and $\bar{\theta}$ of some vertex operators to zero. This results in the inclusion of vertex operators with non-trivial picture numbers. Furthermore, we shall insert PCOs at a finite distance away from the vertex operators with the non-trivial picture numbers. There are in total four different ways to distribute the picture numbers
\begin{equation}\label{eqn:picture dis1}
\langle V_{D}^{(-1,-1)}(z_1,p_1)V_D^{(0,0)}(z_2,p_2)\rangle\,,\quad \langle V_D^{(0,0)}(z_1,p_1)V_D^{(-1,-1)}(z_2,p_2)\rangle\,,
\end{equation}
\begin{equation}\label{eqn:picture dis2}
\langle V_D^{(-1,0)}(z_1,p_1) V_D^{(0,-1)}(z_2,p_2)\rangle\,,\quad \langle V_D^{(0,-1)}(z_1,p_1) V_D^{(-1,0)}(z_2,p_2)\rangle\,.
\end{equation}
Because $p_1^2=p_2^2=0,$ contributions from \eqref{eqn:picture dis1} vanish. Therefore, we will focus on \eqref{eqn:picture dis2} from now on. 

Let us therefore study 
\begin{equation}
\mathfrak{C}'(p_1,p_2):= \langle V_D^{(-1,0)}(z_1,p_1)V_D^{(0,-1)} (z_2,p_2) e^{\phi}T_F (z_0) e^{\bar{\phi}} \bar{T}_F (\bar{z}_0)\rangle\,,
\end{equation}
where $ e^{\phi} T_F(z_0)$ and $e^{\bar{\phi}}\bar{T}_F(\bar{z}_0)$ are due to the insertion of PCOs \cite{Sen:2021tpp,Alexandrov:2021shf}. Note that $T_F=i \sqrt{2/\alpha'}\psi\cdot \partial X+\dots,$ where $\dots$ include the internal degrees of freedom. Because we are in the (odd,odd) spin structure, there are four fermionic zero modes both from left and right moving sectors of the non-compact part of the worldsheet CFT. To have a non-vanishing result, we should therefore soak up the fermionic zero modes by pairing the zero-mode integral with fermion insertions. The component form of the correlator is
\begin{align}
\mathfrak{C}'=& -\frac{2^2\pi^2g_c^2}{\alpha'^2} f_{\mu\nu}^1f_{\rho\sigma}^2\left\langle e^{-\phi}\psi_1^\mu \left(i\bar{\partial}X_1^\nu +\frac{\alpha'}{2}p_1\cdot \bar{\psi}_1\bar{\psi}_1^\nu\right) \left(i\partial X_2^\rho +\frac{\alpha'}{2}p_2\cdot \psi_2\psi_2^\rho\right)e^{-\bar{\phi}}\bar{\psi}^\sigma \right.\nonumber\\
&\qquad\qquad\qquad\times\left(e^{\phi}\psi\cdot\partial X\right)(z_0) \left(e^{\bar{\phi}}\bar{\psi}\cdot\bar{\partial} X\right)(\bar{z}_0) \biggr\rangle \,.
\end{align}
First, we compute
\begin{equation}
\langle \psi_1^\mu p_2\cdot \psi_2\psi_2^\rho \psi^\delta\rangle= p_{2\alpha}\eta(\tau)^4 \epsilon^{\mu\alpha\rho\delta}\,.
\end{equation}
Similarly, we compute
\begin{equation}
\langle p_1 \cdot \bar{\psi}_1\bar{\psi}_1^\nu \bar{\psi}^\sigma \bar{\psi}^{\epsilon}\rangle =p_{1\beta} (\eta(\tau)^4)^*\epsilon^{\beta\nu\sigma\epsilon}\,.
\end{equation}
We pulled out the factor of $(2\pi)^2$ from the fermion correlators into the overall normalization \cite{Kim:2023sfs}. Including this factor $(2\pi)^2$ is crucial, because we are using the convention in which the periodicity of the worldsheet torus is given by $z\sim z+1\sim z+\tau.$ From the $b,c$ ghost system we obtain $|\eta(\tau)|^4,$ and from the $\beta,\gamma$ ghost system with the fermion zero mode saturation we obtain $|\eta(\tau)|^{-4}.$ The boson partition function generates $|\eta(\tau)|^{-8}$ factor. Integral over the closed string momentum along the non-compact directions generates
\begin{equation}
\int \frac{d^4k}{(2\pi)^4}e^{-\pi \tau_2 k^2} =\frac{1}{2^4\pi^4\alpha'^2\tau_2^2}\,.
\end{equation}
The contraction between the bosonic components in the PCOs generate $-\frac{\pi\alpha'}{2\tau_2}.$ The internal part of the CFT produces 
\begin{equation}
\chi:=\text{Tr}_{R,R}^{(int)}\left((-1)^{F+\bar{F}}q^{L_0-\frac{3}{8}}\bar{q}^{\bar{L}_0-\frac{3}{8}}\right)\,.
\end{equation}
Integral over the vertex moduli generates $4\tau_2^2.$ Combining all these, we obtain
\begin{equation}
Z^{(odd,odd)}_\mathcal{T}= \pm\frac{g_c^2}{2\pi \alpha'}\int \frac{d^2\tau}{2\tau_2^2}f_{\mu\nu}^1f_{\rho\sigma}^2 p_{1\alpha}p_{2\beta} \epsilon^{\mu\alpha\rho}\epsilon^{\beta\nu\sigma} \chi\,,
\end{equation}
where the overall $+$ sign is for type IIB and the $-$ sign is for type IIA \cite{Polchinski:1998rr}.

Using the identity
\begin{equation}
\int \frac{d^2\tau}{\tau_2^2}=\frac{\pi}{3}\,,
\end{equation}
we obtain
\begin{equation}
Z^{(odd,odd)}_\mathcal{T}= \pm\frac{1}{3}\frac{g_c^2}{  2^2\alpha' }f_{\mu\nu}^1f_{\rho\sigma}^2 p_{1\alpha}p_{2\beta} \epsilon^{\mu\alpha\rho}\epsilon^{\beta\nu\sigma} \chi\,.
\end{equation}
After simplifying the Levi-Civita symbol, we find
\begin{equation}
f_{\mu\nu}^1f_{\rho\sigma}^2 p_{1\alpha}p_{2\beta}\epsilon^{\mu\alpha\rho}\epsilon^{\beta\nu\sigma}=-2(p_1\cdot p_2)\,.
\end{equation}
Therefore, we arrive at
\begin{equation}
Z^{(odd,odd)}_\mathcal{T}= \mp\frac{1}{3}\frac{g_c^2}{  2\alpha' }p_1\cdot p_2\chi\,.
\end{equation}

As a result, we obtain
\begin{equation}\label{eqn:odd res1}
\mathfrak{G}_{\phi_4,\mathcal{T}}^{(1),(odd)}=\mp\frac{1}{2^4 \cdot 3 \cdot \pi}\chi\,.
\end{equation}
This result agrees with the dimensional reduction of known string loop corrections in 10d string theories and low energy supersymmetry \cite{Antoniadis:2003sw,Alexandrov:2007ec,Liu:2022bfg}.

To compute the contribution from the Klein bottle, one can simply insert the orientation reversal operator $\Omega.$ This is because the Klein bottle partition function is defined as the torus partition function with the orientation reversal operator insertion. This prescription can be also understood as follows. The closed string contribution can be obtained by dimensionally reducing a term in the 10d action that contains $ R\wedge R\wedge R (\partial \phi)^2,$ where $R\wedge R \wedge R$ denotes an appropriate tensor structure of the $R^3$ term. Upon the dimensional reduction on a Calabi-Yau, we have $\int_X R\wedge R\wedge R\sim \chi.$ If on the other hand the target space is an orientifold of a Calabi-Yau, then we expect $\int_{X/\Omega} R\wedge R\wedge R$ to be proportional to the Euler characteristic of the \emph{orientifold}, not the Calabi-Yau itself. Therefore, we argue that $\int_{X/\Omega} R\wedge R\wedge R\sim (\chi+\chi_f)/2,$ where $\chi_f$ is defined as
\begin{equation}
\chi_f:=\text{Tr}_{R,R}\left( (-1)^{F_L+F_R}\Omega q^{L_0-\frac{3}{8}}\bar{q}^{\bar{L}_0-\frac{3}{8}}\right)\,,
\end{equation}
which in geometric phases simplifies to the twisted Euler characteristic \cite{Brunner:2003zm,Denef:2008wq}
\begin{equation}
\chi_f= \sum_{p,q}(-1)^{p+q}(h_+^{p,q}-h_-^{p,q})\,.
\end{equation}
Because in the orientifold the closed string diagram always comes with the projection onto the spectrum that is even under the orientifold action, we need to sum over torus and Klein bottle. As the torus contribution generates a term that is proportional to $\chi,$  we expect that the Klein bottle contribution to be proportional to $\chi_f.$ Therefore, we argue that the following contribution should be generated
\begin{equation}\label{eqn:odd res2}
\mathfrak{G}_{\phi_4,\mathcal{K}}^{(1),(odd)}=\mp\frac{1}{2^4 \cdot 3 \cdot \pi}\chi_f\,.
\end{equation}

As a trivial consistency check, one can consider type I string theory compactified on a Calabi-Yau threefold. Upon dimensionally reducing the term $R\wedge R\wedge R (\partial \phi)^2$ in the 10d action, in the vertex-frame, we obtain a term 
\begin{equation}\label{eqn:type I odd}
\frac{1}{\kappa_4^2}\int \left(-\frac{\chi}{2^4\cdot 3\cdot \pi}\right) (\partial \phi)^2\,.
\end{equation}
As type I string theory can be described as an O5/O9 orientifold compactification of type IIB string theory, the above term is generated by summing over the torus contribution and Klein bottle contribution
\begin{equation}
-\frac{\chi}{2^4\cdot 3\cdot \pi}=\frac{1}{2}\left(\mathfrak{G}_{\phi_4,\mathcal{T}}^{(1),(odd)}+\mathfrak{G}_{\phi_4,\mathcal{K}}^{(1),(odd)}\right)\,,
\end{equation}
note that the factor $1/2$ in the right hand side was introduced due to the sum over topologies. As the torus contribution only generates half of \eqref{eqn:type I odd}, the rest half must come from other diagram, the Klein bottle. In the case of type I string theory, we have $\chi_f=\chi.$ Taking $\chi_f=\chi$ into account, we recover the full result \eqref{eqn:type I odd} by adding the proposed Klein bottle contribution. Note that we have not performed the explicit string amplitude computation for this contribution $\mathfrak{G}_{\phi_4,\mathcal{K}}^{(1),(odd)}.$ It would be interesting to confirm this expectation directly.

\subsection{Full result}
Now, we are ready to combine the amplitudes we computed in this section.

We first recall the contributions from annuli, Mobius strip, and Klein bottle with the even spin structure. At order $\mathcal{O}(\delta),$ we found 
\begin{equation}
\boxed{\mathfrak{G}_{\sigma,\phi}^{(1)}=-\frac{c_\sigma}{2^3\pi^2} \int_0^\infty \frac{dt}{2t^3}\tau_{2\sigma}\left[\text{Tr}_R\left((-1)^{F-\frac{3}{2}}Fq^{L_0-\frac{3}{8}}\right)^{\sigma}_{int}+\frac{3}{2}(n_\sigma^+-n_\sigma^-)\right]\,.}
\end{equation}
Summing over the worlsheet topologies with the even spin structures, we find
\begin{equation}
\mathfrak{G}_{\phi}^{(1),(even)}=\frac{1}{2}\left(\mathfrak{G}_{\mathcal{A},\phi}^{(1),(even)}+\mathfrak{G}_{\mathcal{M},\phi}^{(1),(even)}+\mathfrak{G}_{\mathcal{K},\phi}^{(1),(even)}\right)\,,
\end{equation}

For the odd spin structure, we found that the torus contribution \eqref{eqn:odd res1} and the Klein bottle contribution \eqref{eqn:odd res2} combine to
\begin{equation}\label{eqn:main result}
\boxed{\mathfrak{G}_{\phi}^{(1),(odd)}=\frac{1}{2}(\mathfrak{G}_{\phi_4,\mathcal{T}}^{(1),(odd)}+\mathfrak{G}_{\phi_4,\mathcal{K}}^{(1),(odd)})=\mp\frac{1}{2^5 \cdot 3\cdot \pi} (\chi+\chi_f)\,.}
\end{equation}

Combining all the results we found, we arrive at our main result
\begin{equation}
\boxed{\mathfrak{G}_{\phi}^{(1)}=\mathfrak{G}_{\phi}^{(1),(even)}+\mathfrak{G}_{\phi}^{(1),(odd)}\,.}
\end{equation}

\section{Conclusions}\label{sec:conclusion}
In this paper, we studied the string one-loop correction to the 4d-dilaton kinetic term in 4d $\mathcal{N}=1$ compactifications of string theories. We discuss a few future directions.
\begin{itemize}
	\item In type II string theories, very interestingly, we again found that the string one-loop correction to the 4d dilaton kinetic term in \emph{string-frame} is determined through the new susy index in addition to the Einstein-Hilbert action \cite{Kim:2023sfs} and the one-loop pfaffian of the non-perturbative superpotential \cite{Kim:2023cbh}. Somehow, we are finding that there is a universality of the effective action, in that many of the relevant terms in the effective action at two derivative level are determined by the new susy index of the internal CFT. It would be interesting to understand the reason why the new susy index plays such a prominent role.\footnote{We thank Andreas Schachner for emphasizing this point.}
	\item In this work, we argued for the term $\mathfrak{G}_{\phi_4,\mathcal{K}}^{(1),(odd)},$ but we did not directly compute it. It would be interesting to compute the (odd,odd) spin structure contribution from closed string sectors directly.
	\item In this work, we have not explicitly evaluated the one-loop correction in explicit models. It is extremely important to study the detailed moduli dependence in explicit models.
	\item In this work, we have saturated the RR-tadpole using spacetime-filling D-branes to simplify the computations. In more realistic scenarios, presence of the RR-flux is ubiquitous. Therefore, it would be extremely important to generalize the computation presented in this note to non-trivial flux configurations.
	\item In this work, we have not computed the string one-loop correction to generic moduli fields. It would be important to compute such terms.
	\item It would be interesting to illuminate the role of the Green-Schwarz term in type II string theories. We are actively working on the relation between the Green-Schwarz term and the K\"{a}hler potential in type II string theories \cite{GS}.
\end{itemize}
\section*{Acknowledgements}
The work of MK was supported by a Pappalardo fellowship. MK thanks Daniel Harlow for encouragement. MK thanks Michael Haack, Atakan Hilmi F\i rat, Liam McAllister, Jakob Moritz, Patrick Jefferson, Wati Taylor, Thomas Grimm, Oliver Schlotterer, and Andreas Schachner for discussions. MK thanks Michael Haack for careful reading of this manuscript.
\appendix
\newpage
\section{Conventions}\label{sec:Green function}
In this section, we collect the worldsheet conventions used in this paper. For $\alpha,\beta=0,1,$ we define the Jacobi theta functions
\begin{equation}
\vartheta_{\alpha,\beta}(z|\tau):= \sum_{n\in \Bbb{Z}+\frac{\alpha}{2}} e^{i\pi n \beta}q^{n^2/2}y^n\,,
\end{equation}
where we define
\begin{equation}
q=e^{2\pi i\tau}\,,
\end{equation}
and
\begin{equation}
y=e^{2\pi i z}\,.
\end{equation}
Note that $z$ denotes the flat coordinate of a torus. This shows that we normalized the torus coordinate such that the torus is periodic under $z\sim z+1\sim z+\tau.$ We will oftentimes use a shorthand notation
\begin{equation}
\vartheta_{\alpha,\beta}(\tau):=\vartheta_{\alpha,\beta}(0|\tau)\,.
\end{equation}

We write 
\begin{align}
\vartheta_{0,0}(z|\tau)=&\prod_{n=1}^\infty (1-q^n)\left( 1+(y+y^{-1})q^{n-\frac{1}{2}}+q^{2n-1}\right)\,,\\
\vartheta_{0,1}(z|\tau)=& \prod_{n=1}^\infty (1-q^n)\left( 1-(y+y^{-1})q^{n-\frac{1}{2}}+q^{2n-1}\right)\,,\\
\vartheta_{1,0}(z|\tau)=&q^{\frac{1}{8}}(y^{\frac{1}{2}}+y^{-\frac{1}{2}})\prod_{n=1}^\infty (1-q^n)\left( 1+(y+y^{-1})q^{n}+q^{2n}\right)\,,\\
\vartheta_{1,1}(z|\tau)=&iq^{\frac{1}{8}}(y^{\frac{1}{2}}-y^{-\frac{1}{2}})\prod_{n=1}^\infty (1-q^n)\left( 1-(y+y^{-1})q^{n}+q^{2n}\right)\,.
\end{align}
We define more general Jacobi theta functions using the quasi-periodicity
\begin{align}
\vartheta_{\alpha,\beta}\left(z+\frac{1}{2}\biggr|\tau\right)=&\vartheta_{\alpha,\beta+1}(z|\tau)\,,\\
\vartheta_{\alpha,\beta}\left(z+\frac{\tau}{2}\biggr|\tau\right)=&e^{-i\pi\beta/2}q^{-\frac{1}{8}}y^{-\frac{1}{2}} \vartheta_{\alpha+1,\beta}(z|\tau)\,,\\
\vartheta_{\alpha+2,\beta}(z|\tau)=&\vartheta_{\alpha,\beta}(z|\tau)\,,\\
\vartheta_{\alpha,\beta+2}(z|\tau)=&e^{i\alpha\pi}\vartheta_{\alpha,\beta}(z|\tau)\,.
\end{align}
We define the Dedekind eta function as
\begin{equation}
\eta(\tau):=q^{\frac{1}{24}}\prod_{n=1}^\infty (1-q^n)\,.
\end{equation}

We define the annulus $A,$ M\"{o}bius strip $\mathcal{M},$ and Klein bottle $\mathcal{K}$ by modding out the tori with modulus \cite{Antoniadis:1996vw,Berg:2014ama,Kim:2023sfs}
\begin{equation}
\tau_A=it\,,\qquad \tau_{\mathcal{M}}=\frac{1}{2}+it\,,\qquad \tau_\mathcal{K}=2i t\,,
\end{equation}
by the involutions
\begin{equation}
I_A(z)=1-\bar{z}\,,\qquad I_{\mathcal{M}}(z)=1-\bar{z}\,,\qquad I_{\mathcal{K}}(z)=1-\bar{z}+\frac{\tau_{\mathcal{K}}}{2}\,.
\end{equation}

To compute various correlators on the worldsheet $\sigma,$ we will use the image charge method. The Green's function for free bosonic fields on torus is related to a two point function of bosons as
\begin{equation}
\langle X(z_1)X(z_2)\rangle_\mathcal{T}=G_B (z_1,z_2;\tau)\,,
\end{equation}
where the Green's function is defined as
\begin{equation}
G_B(z_1,z_2;\tau):= -\frac{\alpha'}{2} \log\left|\frac{\vartheta_{1,1}(z_1-z_2|\tau)}{\vartheta_{1,1}'(0|\tau)}\right| +\frac{\pi\alpha'}{\tau_2}(\im (z_1-z_2))^2\,.
\end{equation}
The two point function of two free bosons on $\sigma\neq\mathcal{T}$ is then
\begin{equation}
\langle X(z_1)X(z_2)\rangle_\sigma= \langle X(z_1)X(z_2)\rangle_\mathcal{T}+\langle X(z_1)X(I_\sigma(x_2))\rangle_\mathcal{T}\,.
\end{equation}

Now we collect correlation functions of free fermions. We will mostly follow the conventions of \cite{Berg:2014ama}. We will use a shorthand notation $s$ for the spin structure $(\alpha,\beta).$ The two point functions of free fermions on torus are given by
\begin{equation}
\langle \psi(z_1)\psi(z_2)\rangle_\mathcal{T}^s=G_F(z_1,z_2;\tau,s)\,,\qquad \langle \bar{\psi}(\bar{z}_1)\bar{\psi}(\bar{z}_2)\rangle_\mathcal{T}^s=G_F(z_1,z_2;\tau,s)^*\,,
\end{equation}
and the correlation between holomorphic and anti-holomorphic fermions vanish. Note that we define
\begin{equation}
G_F(z_1,z_2;\tau,s):=\frac{\vartheta_{\alpha,\beta}(z_1-z_2|\tau)\vartheta_{1,1}'(0|\tau)}{\vartheta_{\alpha,\beta}(0|\tau)\vartheta_{1,1}(z_1-z_2|\tau)}\,.
\end{equation}
We summarize the fermionic correlation functions on $\sigma\neq\mathcal{T}$
\begin{align}
\langle \psi(z_1)\psi(z_2)\rangle_\sigma^s=&G_F(z_1,z_2;\tau,s)\,,\\
\langle \psi(z_1)\bar{\psi}(\bar{z}_2)\rangle_\sigma^s=&iG_F(z_1,I_\sigma(z_2);\tau,s)\,,\\
\langle \bar{\psi}(\bar{z}_1)\psi(z_2)\rangle_\sigma^s=&iG_F(I_\sigma (z_1),z_2;\tau,s)\,,\\
\langle \bar{\psi}(\bar{z}_1)\bar{\psi}(\bar{z}_2)\rangle_\sigma^s=&-G_F(I_\sigma(z_1),I_\sigma(z_2);\tau,s)\,.
\end{align}
\section{Genuine eight fermion contractions}\label{app:cancellation}
In this section, we shall explicitly demonstrate the genuine contractions of eight fermions. We have performed the algebraic computations with \texttt{Mathematica}. To simplify the expression, we will define the following shorthand notations
\begin{align}
G_1:=\langle \psi(z_1)\psi(z_2)\rangle_\sigma^s=&G_F(z_1,z_2;\tau,s)\,,\\
G_2:=\langle \psi(z_1)\bar{\psi}(\bar{z}_2)\rangle_\sigma^s=&iG_F(z_1,I_\sigma(z_2);\tau,s)\,,\\
G_3:=\langle \bar{\psi}(\bar{z}_1)\psi(z_2)\rangle_\sigma^s=&iG_F(I_\sigma (z_1),z_2;\tau,s)\,,\\
G_4:=\langle \bar{\psi}(\bar{z}_1)\bar{\psi}(\bar{z}_2)\rangle_\sigma^s=&-G_F(I_\sigma(z_1),I_\sigma(z_2);\tau,s)\,.
\end{align}

Let us recall the eight fermion correlators
\begin{equation}
\mathfrak{C}_8=\frac{\alpha'^2}{4} f^1_{\mu\nu}f^2_{\rho\sigma}\left\langle p_1\cdot\psi_1\psi_1^\mu p_1\cdot \bar{\psi}_1\bar{\psi}_1^\nu p_2\cdot \psi_2 \psi_2^\rho p_2\cdot\bar{\psi}_2\bar{\psi}_2^\sigma\right\rangle\,.
\end{equation}
First, we can start with the contraction
\begin{align}
\mathfrak{K}_8^{(1)\mu\nu\rho\sigma}=&\langle p_1\cdot \psi_1 p_2\cdot \psi_2\rangle \langle \psi_1^\mu \psi_2^\rho \rangle\langle p_1\cdot \bar{\psi}_1p_2\cdot \bar{\psi}_2 \rangle \langle \bar{\psi}_1^\nu\bar{\psi}_2^\sigma\rangle\,\\
=&G_1^2 G_4^2(p_1\cdot p_2)^2 \eta^{\mu\rho}\eta^{\nu\sigma}\,.
\end{align}
By permuting $\{ p_2\cdot \psi_2,\psi_2^\rho,p_2\cdot \bar{\psi}_2,\bar{\psi}_2^\sigma\},$ and correctly taking into the additional signs due to the fermion permutations, we find that there are 24 genuine contractions 
\begin{align}
&G_1^2 G_4^2 (p_1\cdot p_2)^2 \eta ^{\mu\rho } \eta ^{\nu\sigma }\,,
-G_1^2 G_4^2 p_1\cdot p_2 p_2^{\nu } p_1^{\sigma } \eta ^{\mu \rho }\,,
-G_1 G_2 G_3 G_4 p_1\cdot p_2 p_2^{\mu } p_1^{\rho } \eta ^{\nu \sigma }\,,\nonumber\\
&G_1 G_2 G_3 G_4 p_1\cdot p_2 p_2^{\mu } p_1^{\sigma } \eta ^{\nu \rho }\,,
G_1 G_2 G_3 G_4 p_1\cdot p_2 p_2^{\nu } p_1^{\rho } \eta ^{\mu \sigma }\,,
-G_1 G_2 G_3 G_4 (p_1\cdot p_2)^2 \eta ^{\mu \sigma } \eta ^{\nu \rho }\,,\nonumber\\
&-G_1^2 G_4^2 p_1\cdot p_2 p_2^{\mu } p_1^{\rho } \eta ^{\nu \sigma }\,,
G_1^2 G_4^2 p_2^{\mu } p_2^{\nu } p_1^{\rho } p_1^{\sigma }\,,
G_1 G_2 G_3 G_4 p_1\cdot p_2 p_2^{\mu } p_1^{\rho } \eta ^{\nu \sigma }\,,
-G_1 G_2 G_3 G_4 p_2^{\mu } p_2^{\nu } p_1^{\rho } p_1^{\sigma }\,,\nonumber\\
&-G_1 G_2 G_3 G_4 p_1\cdot p_2 p_2^{\nu } p_1^{\rho } \eta ^{\mu\sigma }\,,
G_1 G_2 G_3 G_4 p_1\cdot p_2 p_2^{\nu } p_1^{\rho } \eta ^{\mu\sigma }\,,
G_1 G_2 G_3 G_4 p_1\cdot p_2 p_2^{\mu } p_1^{\rho } \eta ^{\nu \sigma }\,,\nonumber\\
&-G_1 G_2 G_3 G_4 p_1\cdot p_2 p_2^{\mu } p_1^{\sigma } \eta ^{\nu \rho }\,,
-G_1 G_2 G_3 G_4 (p_1\cdot p_2)^2 \eta ^{\mu \rho } \eta ^{\nu \sigma }\,,
G_1 G_2 G_3 G_4 p_1\cdot p_2 p_2^{\nu } p_1^{\sigma } \eta ^{\mu \rho }\,,\nonumber\\
&G_2^2 G_3^2 (p_1\cdot p_2)^2 \eta ^{\mu \sigma } \eta ^{\nu \rho }\,,
-G_2^2 G_3^2 p_1\cdot p_2 p_2^{\nu } p_1^{\rho } \eta ^{\mu \sigma }\,,
-G_1 G_2 G_3 G_4 p_2^{\mu } p_2^{\nu } p_1^{\rho } p_1^{\sigma }\,,
G_1 G_2 G_3 G_4 p_1\cdot p_2 p_2^{\mu } p_1^{\sigma } \eta ^{\nu \rho }\,,\nonumber\\
&G_1 G_2 G_3 G_4 p_1\cdot p_2 p_2^{\nu } p_1^{\sigma } \eta ^{\mu \rho }\,,
-G_1 G_2 G_3 G_4 p_1\cdot p_2 p_2^{\nu } p_1^{\sigma } \eta ^{\mu \rho }\,,
-G_2^2 G_3^2 p_1\cdot p_2 p_2^{\mu } p_1^{\sigma } \eta ^{\nu \rho }\,,
G_2^2 G_3^2 p_2^{\mu } p_2^{\nu } p_1^{\rho } p_1^{\sigma }\,.
\end{align}
By summing over all of the genuine contractions, we find
\begin{align}
\mathfrak{K}_8^{\mu\nu\rho\sigma}=\left(G_2 G_3-G_1 G_4\right)\times \biggl[&G_2 G_3 \left(p_1\cdot p_2 \eta ^{\mu \sigma }-p_2^{\mu } p_1^{\sigma }\right) \left(p_1\cdot p_2 \eta ^{\nu \rho }-p_2^{\nu } p_1^{\rho }\right)\nonumber\\
&-G_1 G_4 \left(p_1\cdot p_2 \eta ^{\mu \rho }-p_2^{\mu } p_1^{\rho }\right) \left(p_1\cdot p_2 \eta ^{\nu \sigma }-p_2^{\nu } p_1^{\sigma }\right)\biggr]\,.
\end{align}

To contract the indices, we will use the following identities
\begin{equation}
f_{\mu\nu}^1f_{\rho\sigma}^2 \eta^{\nu\rho}(p_1\cdot p_2 \eta^{\mu\sigma}-p_2^\mu p_1^\sigma)=0\,,
\end{equation}
and
\begin{equation}
f_{\mu\nu}^1 f_{\rho\sigma}^2 p_1^\rho p_2^\mu \left( p_1\cdot p_2 \eta^{\nu\sigma} -p_2^\nu p_1^\sigma\right)=-2(p_1\cdot p_2)^2\,.
\end{equation}
As a result, we find
\begin{equation}
f_{\mu\nu}^1f_{\rho\sigma}^2\mathfrak{K}_8^{\mu\nu\rho\sigma}=2(p_1\cdot p_2)^2 \left( G_2 G_3-G_1 G_4\right)^2\,.
\end{equation}

\newpage
\bibliographystyle{JHEP}
\bibliography{refs}
\end{document}